\newif\ifsingle
\singletrue 

\newif\ifarxiv
\arxivtrue

\ifsingle
	\documentclass[12pt,onecolumn,journal]{IEEEtran}
\else	
	\documentclass[10pt,twocolumn,journal]{IEEEtran}
\fi

\ifsingle
	 
\else 
	
\fi

\newif\ifnoappc
\noappctrue

\usepackage[top=1in, bottom=0.75in, left=0.75in, right=0.75in]{geometry}
\usepackage{graphicx}
\usepackage{amssymb}
\usepackage{epsfig}
\usepackage{subcaption} 
\usepackage{epstopdf}
\usepackage{algorithm}
\usepackage{algorithmic}
\usepackage{amsmath}
\usepackage{amsfonts}
\usepackage{dsfont}
\usepackage{url}
\usepackage{chemarr}
\usepackage{chemarrow}
\usepackage{bbold}
\usepackage[version=3]{mhchem}
\usepackage{mathabx}
\usepackage{xcolor}
\usepackage{soul}

\newtheorem{remark}{Remark}

\begin{document} 

\title{Designing molecular circuits for approximate maximum a posteriori demodulation of concentration modulated signals} 

\ifarxiv
\author{Chun Tung Chou \\
School of Computer Science and Engineering, University of New South Wales, Sydney, New South Wales 2052, Australia. \\
E-mail: c.t.chou@unsw.edu.au}

\fi

\maketitle

\begin{abstract}
Motivated by the fact that living cells use molecular circuits (i.e. a set of chemical reactions) for information processing, this paper investigates the problem of designing molecular circuits for demodulation. In our earlier work, we use a Markovian approach to derive a demodulator for diffusion-based molecular communication. The demodulation filters take the form of an ordinary differential equation which computes the log-posteriori probability of a transmission symbol being sent. This work considers the realisation of these demodulation filters using molecular circuits assuming the transmission symbols are rectangular pulses of the same duration but different amplitudes, i.e. concentration modulation. This paper makes a number of contributions. First, we use time-scale separation and renewal theory to analytically derive an approximation of the demodulation filter from our earlier work. Second, we present a method to turn this approximation into a molecular circuit. By using simulation, we show that the output of the derived molecular circuit is approximately equal to the log-posteriori probability calculated by the exact demodulation filter if the log-posteriori probability is positive. Third, we demonstrate that a biochemical circuit in yeast behaves similarly to the derived molecular demodulation filter and is therefore a candidate for implementing the derived filter. 

\end{abstract}

\noindent{\bf Keywords:}
Molecular communications; maximum a posteriori; molecular circuits; demodulation ; molecular computation; analog computation

\section{Introduction}
\label{sec:intro} 
Molecular communication is a promising approach to realise communications among nano-bio devices \cite{Akyildiz:2008vt,Nakano:2014fq}. 
In a diffusion-based molecular communication network, transmitters and receivers communicate by using signalling molecules diffusing freely in a fluidic medium. A component in a diffusion-based molecular communication system is the demodulator. The focus of this paper is to realise the demodulator using a molecular circuit, i.e. a set of chemical reactions. 

\begin{figure}
\begin{center}
\ifarxiv
	\includegraphics[trim=0cm 4.5cm 3cm 6.2cm, clip=true, width=1.0\columnwidth]{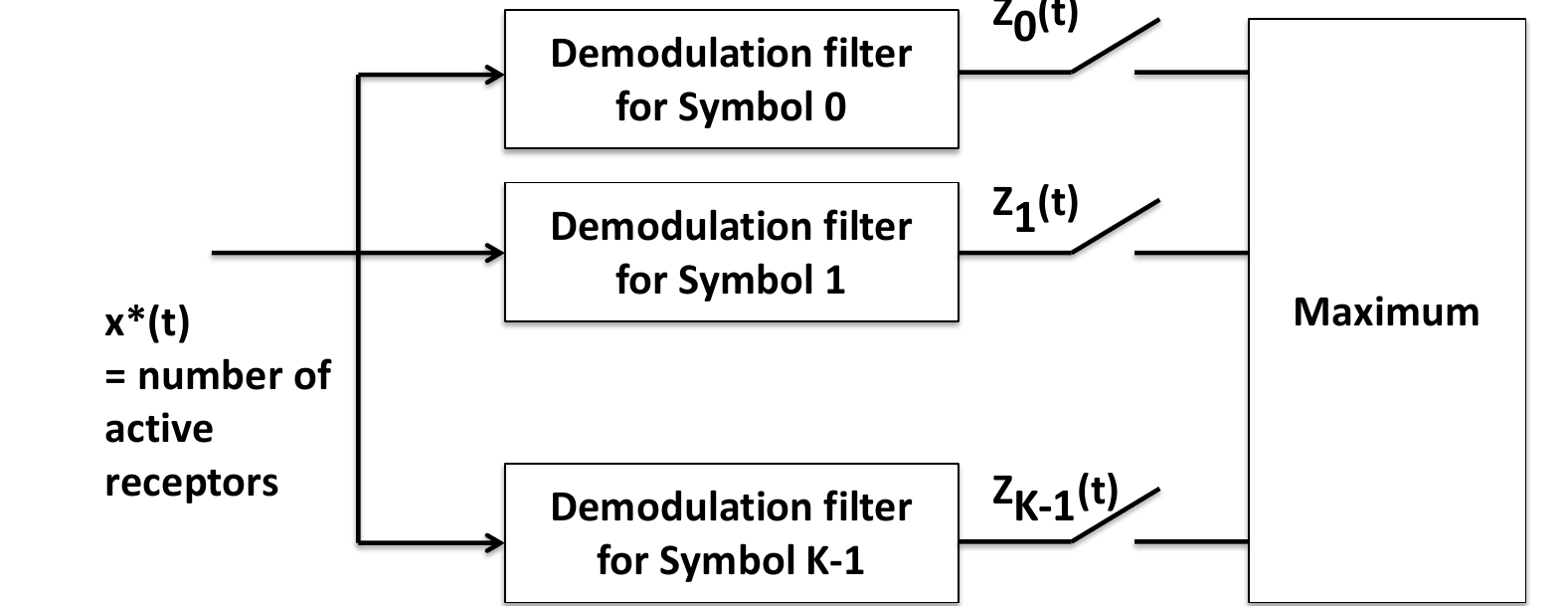}
\fi	
\caption{The demodulator structure.}
\label{fig:demod}
\end{center}
\end{figure}

This paper is built on our earlier work in \cite{Chou:2015ga,Chou:gc} which uses a Markovian approach to design demodulators. The work assumes that the receiver consists of receptors. When the signalling molecules reach the receiver, they can react with these receptors to turn them from inactive to active state. We use the maximum a-posteriori probability (MAP) framework for demodulation. The demodulator consists of a bank of continuous-time demodulation filters, see Fig.~\ref{fig:demod}. The continuous-time input to the demodulators is the number of active receptors. The output of the $k$-th filter $Z_k(t)$ is the log-posteriori probability\footnote{\label{foot:pp} 
Since the signal is continuous-time, the log-posteriori probability diverges. Therefore the log-posteriori probability calculated by these filters is subject to an unknown scaling and constant. However, this is sufficient for demodulation which requires us to determine which transmitted symbol gives, relatively, the largest log-posteriori probability.} of 
$k$-th symbol being sent given the continuous history of receptor activation. A key contribution of \cite{Chou:2015ga} is to derive the ordinary differential equations (ODEs) for these filters. These ODEs describe how to calculate $Z_k(t)$. 

This paper considers the problem of realising the demodulation filters in \cite{Chou:2015ga} using a set of chemical reactions. This is inspired by the fact that chemical reactions are used to decode signals in biological system \cite{Purvis:2013dd}. { Our work is also motivated by recent work \cite{Daniel:2013ke} in synthetic biology to realise computation by using chemicals that are naturally found in living cells. Furthermore, since both chemical based computation and molecular communication are based on molecules, they are therefore compatible. The general goal of our research is to explore how such chemical based computation can be used to realise computation functions for molecular communication because this research may ultimately lead to the implementation of micro- or nano-scale molecular communication devices. Although there is prior work in molecular communication that uses chemical based computation, its emphasis is different. The papers \cite{Unluturk:2015io,Pierobon:2014gl} consider the use of genetic circuits for demodulation but the design of their demodulators is not based on statistics. The papers 
\cite{MrAlessioMarcone:2017te,Marcone:2018kp} present genetic circuits for parity-check while this paper focuses on demodulation. 
 }

Although the demodulation filters in \cite{Chou:2015ga} are described by ODEs and chemical reactions can also be modelled by ODEs, there are two difficulties to realise these demodulation filters. First, these filters require some computation, e.g. the computing of derivative, which is complex to implement by using chemical reactions. Second, these filters need an internal model of the expected signal; therefore, a question is how this internal model can be encoded. In this paper, we show how we can overcome these two difficulties for the case where the transmission symbols are rectangular pulses of the same duration but different amplitudes, i.e. concentration modulation (CM). {\color{black} Although the solution to these two difficulties allows us to find a way to realise the demodulator using chemical reactions in general, it does not address the problem of identifying the chemical species that can actually implement the demodulator. By using the experimental data in \cite{Hansen:2013fs}, we show that a naturally found biochemical circuit in yeast can be used to implement the demodulation filters.} This paper makes the following contributions:
\begin{itemize} 
\item By using time-scale separation \cite{khalil2014nonlinear} and renewal theory \cite{Grimmett}, we analytically derive an approximation of the demodulation filters from our earlier work. This approximation removes some of the computation which is complex to implement by using chemical reactions.  
\item We derive a method to turn the approximation into a molecular circuit. This is done by deriving a method to encode the amplitude of the transmission symbol into the molecular circuit. 
\item By using simulation, we show that the output of this molecular circuit is approximately equal to the log-posteriori probability calculated by the demodulation filter if the log-posteriori probability is positive. 
\item We show that a biochemical circuit in yeast behaves similarly to the derived molecular circuit. This biochemical circuit is therefore a candidate to implement the derived molecular circuit. 
\item For the case where the transmitter uses two symbols, we propose a chemical reaction to approximate the maximum block in the demodulator in Fig.~\ref{fig:demod}.
\end{itemize}

The rest of the paper is organised as follows. Sec.~\ref{sec:related} discusses related work. We then present the modelling assumptions and background work in Sec.~\ref{sec:bg}. In Sec.~\ref{sec:simp}, we assume diffusion is absent and derive a molecular circuit that approximately realises the demodulation filter. We then demonstrate in Sec.~\ref{sec:promotor} that a biochemical circuit in yeast behaves similarly to the derived molecular circuit. In Sec.~\ref{sec:full} we show how the molecular circuit derived in  Sec.~\ref{sec:simp} can be adapted to diffusion-based molecular communication. Finally, Sec.~\ref{sec:con} concludes the paper. 

\section{Related Work} 
\label{sec:related} 
For a recent survey on molecular communication, see \cite{Farsad:2016eu}. Although there was much earlier work on demodulation, see e.g. \cite{Noel:2014hu,Mahfuz:2014vs,ShahMohammadian:2013jm}, this paper differs from the earlier work in two key aspects. First, most earlier work assumed that the demodulation is based on one sample point per symbol; however, this work assumes that demodulation is based on the continuous history of the number of active receptors. Second, most earlier work did not consider a demodulator which is made entirely from chemical reactions. 

Our assumption of using the continuous history of the number of active receptors for demodulation leads to a demodulator which uses analog filters. The use of analog filters for demodulation was studied in \cite{Chou:2015ga,Chou:gc,Awan:2017fm}, but no molecular circuit realisation was provided. A recent work \cite{Kuscu:2018iy} presented two different molecular circuits for demodulation but their circuits used one sample point per symbol for demodulation rather than continuous history. 

There were other examples of using molecular circuits for molecular communication. The paper \cite{Pierobon:2014gl} presented a biological circuit for molecular communication from a system-theoretic perspective. There was also work on using analog circuits for soft detection \cite{MrAlessioMarcone:2017ue} and parity check decoder \cite{MrAlessioMarcone:2017te}. The key difference between these few pieces of work and ours is that they use one sample point per symbol. Lastly, there is also work on using chemical reactions to produce transmission signals for molecular communication, see \cite{Deng:2017uw}. 

The use of chemical reactions to implement analog computation is an active area of research in molecular computing and synthetic biology, see \cite{Daniel:2013ke,Oishi:2011ig,Salehi:fz,Song:2016cu}. However, the problem of using chemical reactions to implement an analog filter based demodulator does not seem to have been done before. 

Note that part of Secs.~\ref{sec:simp} and \ref{sec:full} appeared in an early conference publication \cite{Chou2018:nanocom}. In particular, the molecular circuit in Sec.~\ref{sec:simp} is slightly different from that in \cite{Chou2018:nanocom}; this is so that we can relate the molecular circuit in Sec.~\ref{sec:simp} to the biochemical circuit in Sec.~\ref{sec:promotor}. 

\section{Model and Background} 
\label{sec:def}
\label{sec:bg} 
This section first presents the set up of our molecular communication system and then summarises our earlier results on optimal demodulation \cite{Chou:2015ga} using analog filters. We assume that there is no inter-symbol interference (ISI) and we focus on the demodulation of one transmission symbol. There is no loss in generality because the same demodulator will be used whether ISI is present or not. The reader can refer to our earlier work \cite{Chou:2015ga} on how ISI can be handled using decision feedback for ligand-receptor circuit when the number of receptors is large\footnote{\color{black} Note that the ISI handling algorithm in \cite{Chou:2015ga} is not chemical based and the design of such algorithm requires much further work. One problem to resolve is that the value of demodulator at the end of a symbol duration may be non-zero and a method is needed to reset it to zero for the next symbol duration. Another problem to resolve is how to implement the algorithm logic and state using chemical based computation. Methods may also be needed to prevent the demodulator from getting saturated. We will leave it to further work.}.  

The modelling framework of this paper mostly follows our previous work \cite{Chou:2015ga} but uses the receptor model from \cite{Chou:gc} to make the problem more tractable. We model the medium as a rectangular prism and divide the medium into voxels. We assume that the transmitter and the receiver each occupies one voxel\footnote{\color{black} Note that it may possible to use the framework in \cite{Awan:2017fm} to deal with the case where a receiver consists of multiple voxels. This is work in progress.}. Although it may not be physically realistic for the receiver to have a cubic shape, this simplified geometry allows us to focus on the signal processing aspect of the receiver.  

The transmitter communicates with the receiver using one type of signalling molecule \cee{S}. The transmitter uses $K$ different symbols where each symbol is characterised by a time-varying emission rate of signalling molecules. We will index the transmission symbols by using $k$ where $k = 0, ..., K-1$. Once the signalling molecules have been emitted into the transmitter voxel, they are free to diffuse in the medium. 

The receiver consists of receptors. These receptors can exist in two states: inactive state \cee{X} and active state \cee{X_*}. An inactive receptor can be activated by a signalling molecule. The activation and deactivation reactions are, respectively: 
\begin{subequations}
\label{cr:all} 
\begin{align}
\cee{
S + X &  ->[g_+] S + X_* \label{cr:on}  \\
X_* &  ->[g_-] X* \label{cr:off}}
\end{align}
\end{subequations}
where $g_+$ and $g_-$ are propensity function constants \cite{Erban:2007we}. Let $x(t)$ and $x_*(t)$ denote, respectively, the {\sl number} of \cee{X} and \cee{X_*} molecules at time $t$. Note that both $x(t)$ and $x_*(t)$ are piecewise constant because they are molecular counts. We assume \cee{X} and \cee{X_*} can only be found in the receiver voxel and are uniformly distributed within the voxel. We further assume $x(t) + x_*(t)$ is a constant for all $t$ and we denote this constant by $M$. 

We model the dynamics of diffusion and chemical reactions by using reaction-diffusion master equation (RDME)\footnote{\color{black} There are three major classes of stochastic models for modelling systems with both diffusion and reactions. They are the Smoluchowski equation, RDME and the Langevin equation \cite{Erban:2007we}. The Smoluchowski equation is based on particle dynamics. It is a fine grained model but hard to work with analytically. Both RDME and Langevin are easier to work with analytically but master equation has a finer scale and granularity compared to the Langevin equation \cite{DelVecchio:book}. Therefore we choose to use RDME which allows us to use Markovian theory for analysis and is at the same time a finer grained model.} \cite{Gardiner}. This means that $x_*(t)$ is a realisation of a continuous-time Markov chain. 

In the formulation of the demodulation problem, we will assume that at time $t$, the data available to the demodulation problem are $x_*(\tau)$ for all $\tau \in [0,t]$; in other words, the data are continuous in time and are the history of the counts of \cee{X_*} up to time $t$. We will use ${\cal X}_*(t)$ to denote the continuous-time history of $x_*(t)$ up to time $t$. 

We adopt a MAP framework for detection. Let ${\mathbf P}[k | {\cal X_*}(t)]$ denote the posteriori probability that symbol $k$ has been sent given the history ${\cal X_*}(t)$. 
Instead of working with ${\mathbf P}[k | {\cal X_*}(t)]$, we will work with its logarithm. Let $L_k(t) = \log ({\mathbf P}[k | {\cal X_*}(t)])$. 

As discussed in Footnote \ref{foot:pp} on Page \pageref{foot:pp}, the log posteriori probability $L_k(t)$ diverges but we are able to determine a scaled and shifted value of $L_k(t)$. In order to simplify the terminology, we will refer to the scaled and shifted version of the log posteriori probability simply as log posteriori probability and will denote it by $L_k(t)$. In \cite{Chou:2015ga}, we show that $L_k(t)$ obeys the following ODE:
\begin{align}
\frac{dL_k(t)}{dt} =& \left[ \frac{dx_*(t)}{dt} \right]_+  \log( {\mathbf E}[n_R(t) | k, {\cal X_*}(t)] ) - \nonumber \\ 
& g_+ (M - x_*(t)) {\mathbf E}[n_R(t) | k, {\cal X_*}(t)] 
\label{eqn:logpp_dd} 
\end{align}
where $n_R(t)$ is the number of signalling molecules in the receiver voxel and $[\xi]_+ = \max(\xi,0)$. The term ${\mathbf E}[n_R(t) | k, {\cal X_*}(t)]$ in Eq.~\eqref{eqn:logpp_dd} is the prediction of the mean number of signalling molecules in the receiver voxel using the history of receptor state ${\cal X_*}(t)$ and can be obtained by solving an optimal Bayesian filtering problem. The initial value $L_k(0)$ is the logarithm of the prior probability that Symbol $k$ is being sent. Since $x_*(t)$ is a piecewise constant signal counting the number of \cee{X_*} molecules, its derivative is a sequence of Dirac deltas at the time instants that \cee{X} is activated or \cee{X_*} is deactivated. Note that the Dirac deltas corresponding to the activation of \cee{X} carries a positive sign and the $[  \; ]_+$ operator keeps only these. 

{\color{black} Eq.~\eqref{eqn:logpp_dd} shows what information within the history ${\cal X_*}(t)$ contributes towards the log posteriori probability computation. Note that the first term on the RHS of Eq.~\eqref{eqn:logpp_dd} is non-zero only at the time instants that an \cee{X} molecule is activated. This shows that the activation of \cee{X} contains information. Recall that $x(t) + x_*(t) = M, \forall \, t$, therefore the factor $(M - x_*(t))$ in the second term equals to $x(t)$. This shows that information is contained in the number of inactive \cee{X} molecules over time.} 

Note that it is difficult to use Eq.~\eqref{eqn:logpp_dd} for computation. This is because the computation of ${\mathbf E}[n_R(t) | k, {\cal X_*}(t)]$ in Eq.~\eqref{eqn:logpp_dd} is an optimal filtering problem which requires extensive computation. In \cite{Chou:2015ga}, we proposed to overcome this problem by using prior knowledge. If Symbol $k$ is transmitted, the mean number of signalling molecules in the receiver voxel is ${\mathbf E}[n_R(t) | k]$ and we denote this by $\sigma_k(t)$. We assume that the receiver uses $\sigma_k(t)$ as an internal model for demodulation. The use of internal models is fairly common in signal processing and communication, e.g. a matched filter. 
After replacing ${\mathbf E}[n_R(t) | k, {\cal X_*}(t)]$ in Eq.~\eqref{eqn:logpp_dd} by $\sigma_k(t)$, we arrive at the following demodulation filter: 
\begin{align}
\frac{dZ_k(t)}{dt} =  \left[ \frac{dx_*(t)}{dt} \right]_+ \log(\sigma_k(t)) - g_+ (M - x_*(t)) \sigma_k(t)
\label{eqn:logmap_s}
\end{align}
where $Z_k(0)$ is initialised to the logarithm of the prior probability that the transmitter sends Symbol $k$. If the demodulator makes the decision at time $t$, then the demodulator decides that Symbol $\hat{k}$ has been transmitted if $\hat{k} = {\arg\max}_{k = 0, ..., K-1} Z_k(t)$.  

Fig.~\ref{fig:demod} shows the demodulator structure where Eq.~\eqref{eqn:logmap_s} is used as demodulation filters. Although Eq.~\eqref{eqn:logpp_dd} is the {\sl optimal} demodulation filter, the replacement of ${\mathbf E}[n_R(t) | k, {\cal B}(t)]$ by $\sigma_k(t)$ makes the demodulation filter in Eq.~\eqref{eqn:logmap_s} {\sl sub-optimal}. Numerical experiments in \cite{Chou:2015ga} show that Eq.~\eqref{eqn:logmap_s} approximates Eq.~\eqref{eqn:logpp_dd} well. In contrast to Eq.~\eqref{eqn:logpp_dd}, the demodulation filter in Eq.~\eqref{eqn:logmap_s} makes use of the instantaneous value of $x_*(t)$ and does not required past history of $x_*(t)$ to be stored.


Although the ODE \eqref{eqn:logmap_s} can be readily solved numerically by a modern computer using $x_*(t)$ as the input, an interesting problem is whether we can realise the computation using a set of chemical reactions. Although much progress has been made in recent years in synthetic biology to realise synthetic analog computation, see e.g. \cite{Daniel:2013ke,Soloveichik:2010be,Oishi:2011ig}, it is still an open problem as to whether an {\sl exact} realisation of Eq.~\eqref{eqn:logmap_s} is possible. The difficulty lies with computing the derivative $\left[ \frac{dx_*(t)}{dt} \right]_+$ and how to introduce prior knowledge $\sigma_k(t)$ in Eq.~\eqref{eqn:logmap_s}. In this paper, we will choose the waveform of the transmission symbols so that we can approximately realise Eq.~\eqref{eqn:logmap_s} by using chemical reactions. 


\section{Simplified demodulation problem} 
\label{sec:simp} 
In this section, we will study a simplified demodulation problem to gain some insight into the properties of the demodulation filter \eqref{eqn:logmap_s}. We will then use this insight in Sec.~\ref{sec:full} to realise the demodulation filter \eqref{eqn:logmap_s}.  

We will assume in this section that the transmitter and the receiver are co-located in the same small volume. This allows us to ignore diffusion. (We will add diffusion back in Sec.~\ref{sec:full}.) For this section, we assume that the transmitter can precisely manipulate the number of signalling molecules in this small volume. If the transmitter sends Symbol $k$, then the number of signalling molecules in the small volume at time $t$ is a deterministic signal $\lambda_k(t)$. We will use $u(t)$ to denote the signal that is sent by the transmitter where $u(t)$ is one of $\lambda_k(t)$'s. 

As in Sec.~\ref{sec:def}, the receiver consists of receptors defined by the reactions in \eqref{cr:all}. We model the reaction dynamics by using chemical master equation \cite{Gardiner}. As a result, the number of active receptors $x_*(t)$ is a realisation of a continuous-time Markov chain. Due to stochastic chemical reactions, a deterministic transmission by the transmitter can result in different $x_*(t)$. We again consider ${\cal X}_*(t)$ as the observation and adopt a MAP framework for demodulation. Let $L_k(t)$ be the log-posteriori probability up to an unknown shift. It can be shown that the {\sl optimal} demodulation filters are:
\begin{align}
\frac{dL_k(t)}{dt} =  \left[ \frac{dx_*(t)}{dt} \right]_+ \log(\lambda_k(t)) - g_+ (M - x_*(t)) \lambda_k(t)
\label{eqn:logmap_simp}
\end{align}
Eq.~\eqref{eqn:logmap_simp} can be derived in the same way as Eq.~\eqref{eqn:logpp_dd}. Intuitively, we can consider Eq.~\eqref{eqn:logmap_simp} as a special case of Eq.~\eqref{eqn:logpp_dd} with ${\mathbf E}[n_R(t) | k, {\cal X_*}(t)]$ replaced by $\lambda_k(t)$ because the transmitter signal $\lambda_k(t)$ is deterministic and is co-located with the receiver. Note that Eqs.~\eqref{eqn:logmap_simp}, \eqref{eqn:logpp_dd} and \eqref{eqn:logmap_s} have the same form. 

Our goal is to realise Eq.~\eqref{eqn:logmap_simp} approximately by using chemical reactions. We will first derive an intermediate approximation of Eq.~\eqref{eqn:logmap_simp} in Sec.~\ref{sec:sep}. This intermediate approximation can be interpreted as a matched filter and this is discussed in Sec.~\ref{sec:matching}. We then present a molecular realisation of this approximation in Sec.~\ref{sec:mol0}. 

\subsection{An intermediate approximation of Eq.~\eqref{eqn:logmap_simp}} 
\label{sec:sep}
We will make use of time-scale separation, which is a property found in many biological systems \cite{DelVecchio:2016iv}, to derive an approximation of Eq.~\eqref{eqn:logmap_simp}. For this paper, we take time-scale separation to mean a mixture of fast and slow dynamics. Specifically, we will assume that the input $u(t)$ is slow {\sl relative} to the speed of the activation and deactivation reactions \eqref{cr:all}. We first present an intuitive argument to explain the rationale of this assumption. {\color{black} We consider the reactions in \eqref{cr:all} as a dynamical system where the input $u(t)$ is the counts of \cee{S} molecules over time and the output is the counts of \cee{X_*} over time. This dynamical system can be modelled by a set of nonlinear ODEs. This set of ODEs can be linearised and the resulting dynamical system behaves as a first-order low-pass filter. The pass band of this low-pass filter depends on the speed of the chemical reactions in \eqref{cr:all}. If the reactions are fast (resp. slow), then the low-pass filter has a wider (narrower) pass band. (Note: The frequency domain properties of chemical reactions were studied in earlier molecular communication literature, see e.g.~\cite{Pierobon:2014gl,Chou:2013jd}.) In the following argument, we will assume the input signal $u(t)$ is fixed and has it energy concentrated in the frequency band up to, say $B_u$ Hz. We consider the following two cases: (i) Reactions are slow relative to the input; (ii) Reactions are fast relative to the input. For case (i), slow reactions means the pass band is narrower than $B_u$. This means that the reactions will filter out the high frequency components in the input signal. This results in information loss and is not conducive for detection. For case (ii), fast reactions means the pass band is wider than $B_u$. This means that the information in the input is mostly preserved.} Hence, we will assume that the input $u(t)$ is slow relative to the reactions \eqref{cr:all}.  

In order to analytically derive an approximation of Eq.~\eqref{eqn:logmap_simp}, we will choose the transmitter symbol $\lambda_k(t)$ as a rectangular pulse of duration $d$. The time profile for Symbol $k$ is: $\lambda_k(t) = a_k$ for $0 \leq t < d$ and $\lambda_k(t) = b$ for $t \geq d$, where $a_k$ and $b$ are, respectively, the amplitude of the pulse when it is ON and OFF. We further assume $a_k \gg b \geq 1, \forall k$. Note that we need a positive $b$ to ensure $\log(\lambda_k(t))$ in Eq.~\eqref{eqn:logmap_simp} is well-defined. Since these symbols have the same duration but different amplitudes, they define CM. Since the input $u(t)$ is one of $\lambda_k(t)$'s, the input $u(t)$ has the form: In the time interval $t \in [0,d)$, $u(t) = a$ where $a$ is one of $a_k$'s, and for $t \geq d$, $u(t) = b$. 

A key goal of deriving the intermediate approximation is to find a way to approximate the first term in Eq.~\eqref{eqn:logmap_simp} because it is difficult to use chemical reactions to compute derivatives. The contribution of the first term on the right-hand side (RHS) of Eq.~\eqref{eqn:logmap_simp} to $L_k(t)$ can be written as $L_{1k}(t) = \int_0^t \left[ \frac{dx_*(\tau)}{d\tau} \right]_+ \log(\lambda_k(\tau)) d\tau$. Assuming $t < d$, then $L_{1k}(t)$ equals to $\log(a_k)$ times the total number of times that the receptors have been activated in the time interval $[0,t]$. It can be shown that the mean time between two consecutive activations of a receptor is $\frac{1}{g_+ a} + \frac{1}{g_-}$. If the duration $d$ and amplitude $a$ are chosen such that $d \gg \frac{1}{g_+ a} + \frac{1}{g_-}$ (which is the time-scale separation assumption), then there are going to be many activations of the receptors when the pulse is ON. This allows us to use the renewal theorem \cite{Grimmett} to approximate the integral $L_{1k}(t)$. In Appendix \ref{sec:no_deriv}, we derive the following ODE: 
\begin{eqnarray}
\frac{d\hat{L}_k(t)}{dt} =  g_- \; x_*(t) \times \left\{ \log(\lambda_k(t)) - \frac{\lambda_k(t)}{u(t)} \right\}.  
\label{eq:Lhat}
\end{eqnarray}
and show that $\hat{L}_k(t) \approx L_k(t)$ in the time interval $t \in [0,d)$. We will call $\hat{L}_k(t)$ the intermediate approximation. 

\begin{figure*}[t]
    \centering
    \begin{subfigure}[t]{0.45\textwidth}
        \centering
        \includegraphics[scale=0.40]{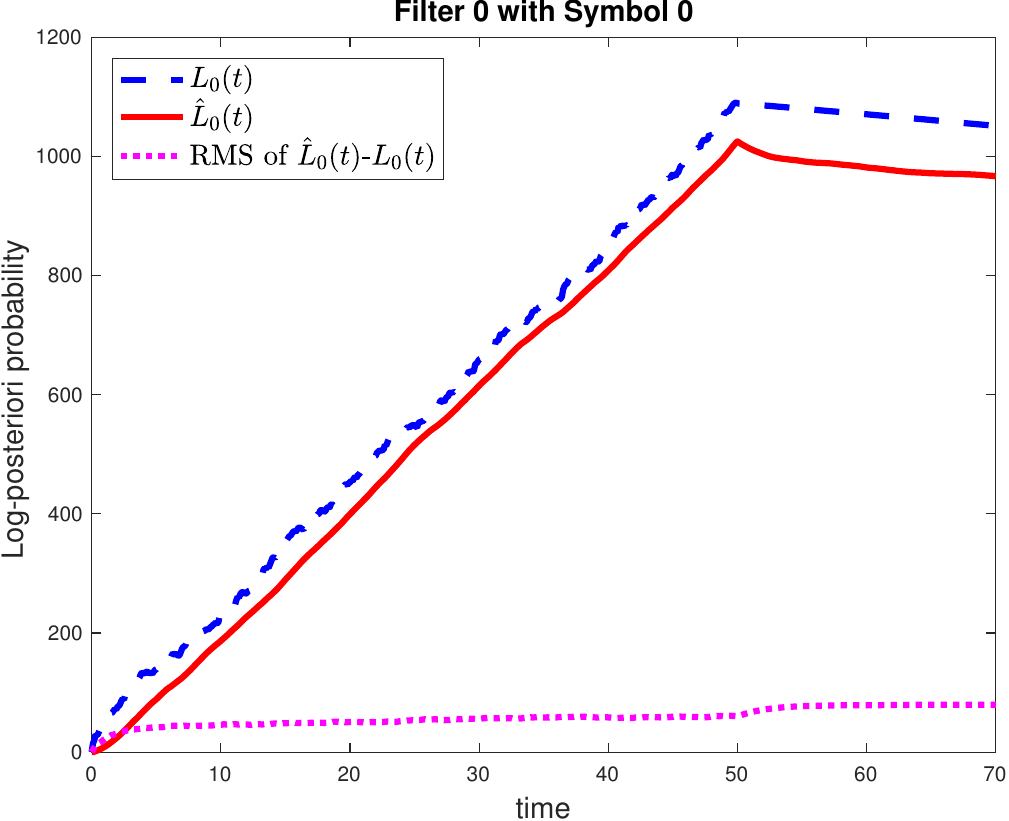}
        \caption{Filter 0 with Symbol 0.}
        \label{fig:local_1_00}
    \end{subfigure} 
    \begin{subfigure}[t]{0.45\textwidth}
        \centering
        \includegraphics[scale=0.40]{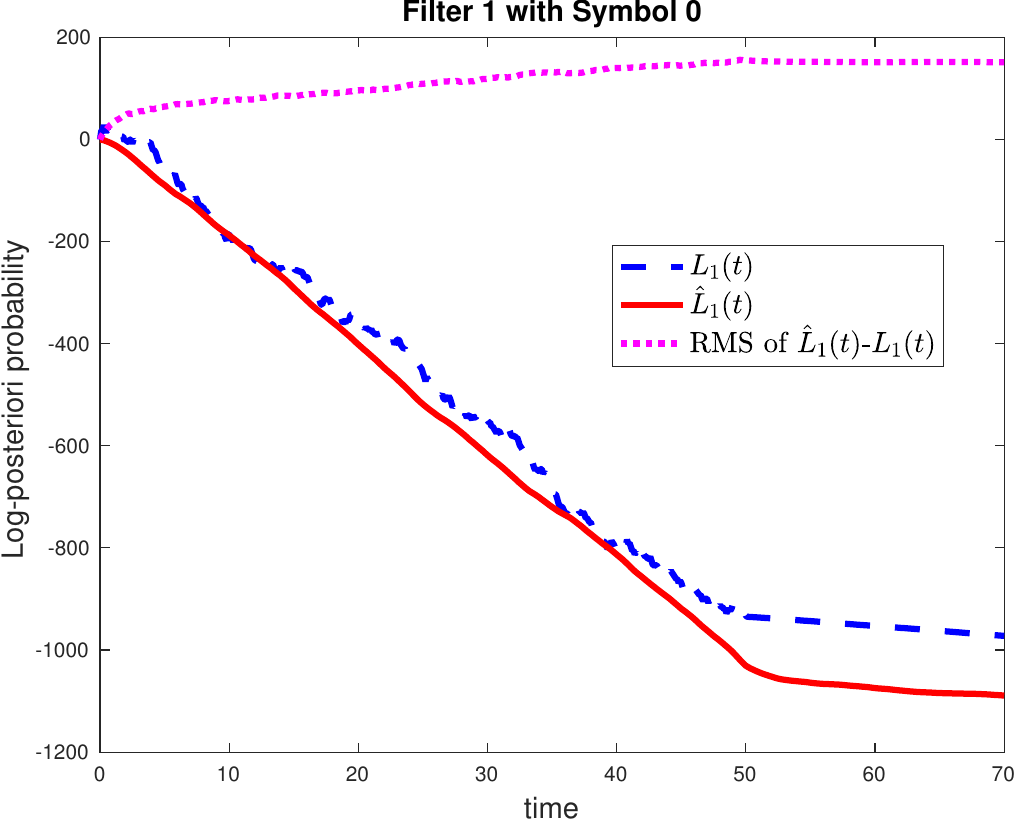}
        \caption{Filter 1 with Symbol 0.}
        \label{fig:local_1_01}
    \end{subfigure}   
   
    \begin{subfigure}[t]{0.45\textwidth}
        \centering
        \includegraphics[scale=0.40]{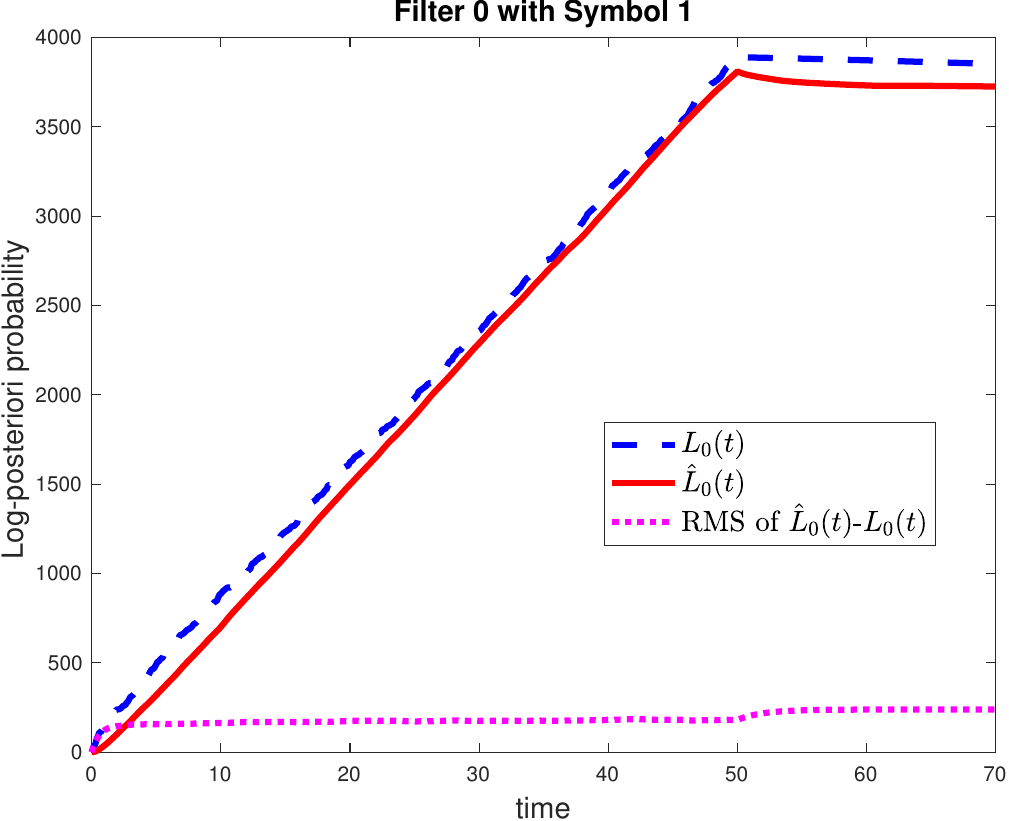}
        \caption{Filter 0 with Symbol 1.}
        \label{fig:local_1_10}
    \end{subfigure} 
    \begin{subfigure}[t]{0.45\textwidth}
        \centering
        \includegraphics[scale=0.40]{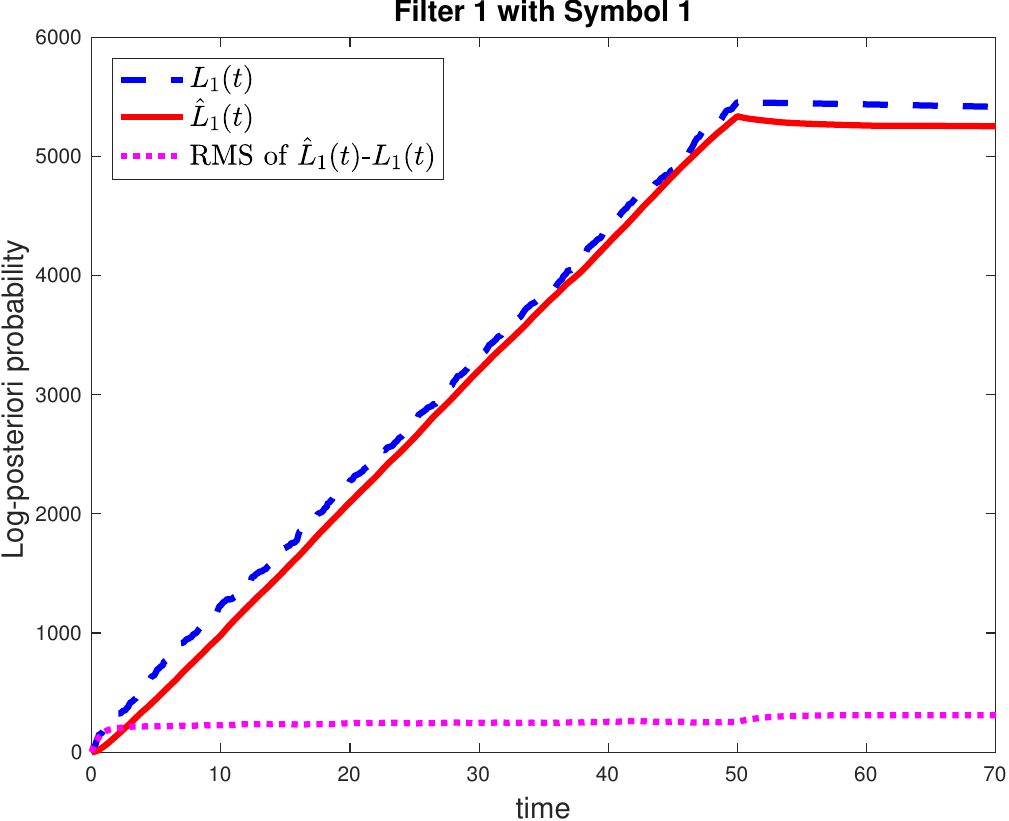}
        \caption{Filter 1 with Symbol 1.}
        \label{fig:local_1_11}
    \end{subfigure}    
\caption{Comparing $L_k(t)$, $\hat{L}_k(t)$ and the RMS of $L_k(t)-\hat{L}_k(t)$.}
\label{fig:local_1}
\end{figure*}

We present a numerical example to show the properties of the intermediate approximation $\hat{L}_k(t)$. This example assumes $K = 2$, $a_0 = 11$, $a_1 = 58$, $d = 50$, $b = 1$, $g_+ = 0.02$, $g_- = 0.5$ and $M = 100$. We use the Stochastic Simulation Algorithm (SSA) \cite{Gillespie:1977ww} to obtain two realisations of $x_*(t)$, one for the case where the input is $\lambda_0(t)$ (Symbol 0) and the other for $\lambda_1(t)$ (Symbol 1). We then use the $x_*(t)$ obtained for Symbol 0 with Eq.~\eqref{eqn:logmap_simp} to compute $L_k(t)$ for $k = 0, 1$, and with Eq.~\eqref{eq:Lhat} to compute $\hat{L}_k(t)$ for $k = 0, 1$; similarly for Symbol 1. Fig.~\ref{fig:local_1} compares $L_k(t)$ and $\hat{L}_k(t)$ for the 4 combinations of input symbols and demodulation filters. It can be seen that for time $t < 50 \; (= d)$, $\hat{L}_k(t)$ (the red solid line) approximates $L_k(t)$ (the blue dashed lines) well. We repeat the above numerical experiment 100 times with independent SSA simulations. We calculate the root-mean-square (RMS) error between $L_k(t)$ and $\hat{L}_k(t)$. Fig.~\ref{fig:local_1} shows that the RMS errors (the magenta dotted lines) are small. 

Having shown that the intermediate approximation $\hat{L}_k(t)$ is a good approximation of the true posteriori probability $L_k(t)$ for $t \leq d$, we will now discuss demodulation. Continuing on the numerical example, if the demodulation decision is to be made at time $t = d$, i.e. at the end of the pulse, then we can see from Fig.~\ref{fig:local_1} that if Symbol $i$ is used as the input, then the intermediate approximation $\hat{L}_i(d)$ has a higher value, hence correct demodulation. 

For $t \geq d$, we see from Fig.~\ref{fig:local_1} that the quality of approximation is poor because the time-scale separation does not hold. Note that this poor approximation will not be an issue because we will see in Sec.~\ref{sec:mol0} that we will not be using $x_*(t)$ for $t \geq d$ in computing log-probability. This is understandable because our CM symbols have different amplitudes in $t \in [0,d)$ but the same amplitude for $t \geq d$; this means there is little information on the symbol being sent in $x_*(t)$ for $t \geq d$. 

Note that there are still issues with using chemical reactions to realise the intermediate approximation and we will address them in Sec.~\ref{sec:mol0}. Before that, we will provide further insight into the intermediate approximation and show that it can be interpreted as a matched filter. 

\subsection{Matched filter interpretation} 
\label{sec:matching}

For $ t < d$, we can write the intermediate approximation $\hat{L}_k(t)$ as:
\begin{eqnarray}
\hat{L}_k(t)=  \left\{ \log(a_k) - \frac{a_k}{a} \right\}  \times  \int_0^t  g_- x_*(\tau)  d\tau  \mbox{ for } t < d 
\label{eq:Lhat_1}
\end{eqnarray}
where we have used the facts that $\lambda_k(t) = a_k$ and $u(t) = a$ for $t < d$. The integral in Eq.~\eqref{eq:Lhat_1} is common to all demodulation filters and is independent of $k$, therefore it does not play a role in distinguishing between different transmission symbols. On the other hand, the factor within the curly brackets in Eq.~\eqref{eq:Lhat_1} depends on $\lambda_k(t)$ and $u(t)$ only; therefore this factor holds the key to understanding how the intermediate approximation differentiates the different transmitter symbols. 

The demodulator makes the decision by choosing the $k$ that maximises $\hat{L}_k(t)$, so this is the same as choosing the $k$ that maximises $\left\{ \log(a_k) - \frac{a_k}{a} \right\}$. Consider the function $phi_a(z) =  \left\{  \log(z) - \frac{z}{a}  \right\}$
which has the same form as the front factor in Eq.~\eqref{eq:Lhat_1} and is parameterised by $a$. In the function $\phi_a(z)$, $a$ and $z$ play the roles of the amplitude of, respectively, the input signal and the reference signal.  It can be shown that $\phi_a(z)$ is concave and has a unique maximum at $z = a$. This means that for a given input amplitude $a$, the amplitude of the reference signal that maximises $\phi_a(z)$ is $z = a$, which means correct demodulation. This explains how matched filtering is performed for CM signals. 
\ifarxiv
Continuing from the earlier example, Fig.~\ref{fig:matching} plots $\phi_{a=11}(z)$ and $\phi_{a=58}(z)$ for the two transmission symbols. We can see that these functions peak when the input amplitude matches the reference amplitude.  
\begin{figure}[t]
    \centering
        \includegraphics[scale=0.40]{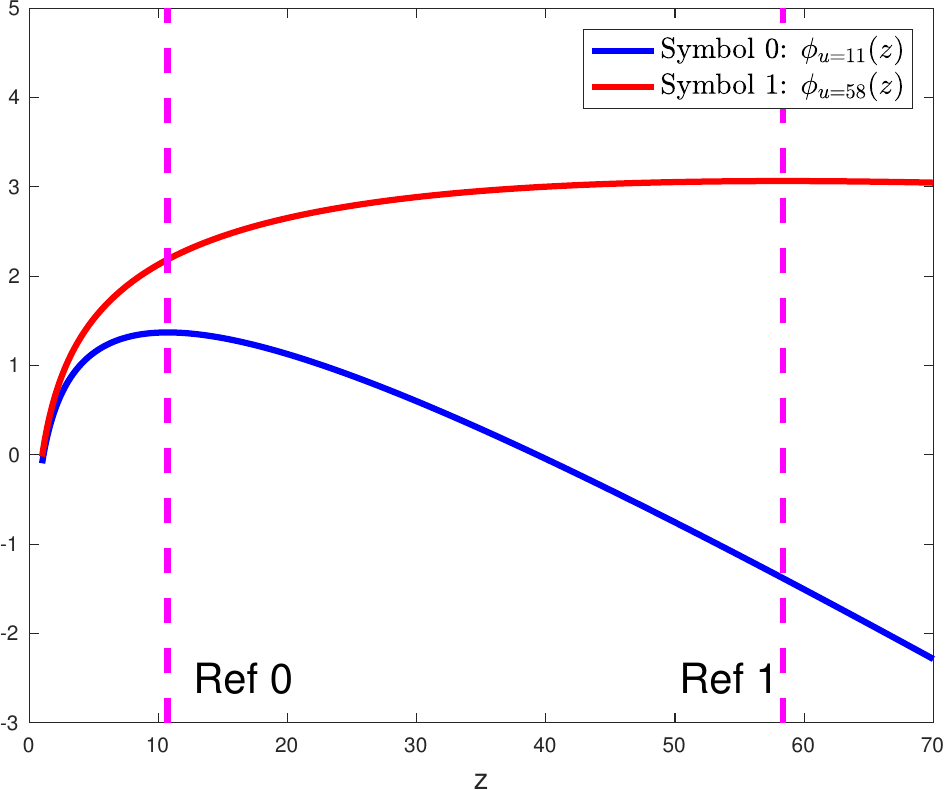}
        \caption{Plotting $\phi_{a = 11}(z)$ and $\phi_{a = 58}(z)$.}
        \label{fig:matching}
\end{figure}
\fi


\subsection{Molecular Realisation of the CM Demodulator} 
\label{sec:mol0} 
We have now shown that the intermediate approximation can be used to demodulate CM inputs, we now investigate how we can realise this CM demodulator using chemical reactions. There are a few of modifications that we need to make. 

First we need to know that log-probability can take any real value, but chemical concentration can only be non-negative. Although \cite{Oishi:2011ig} has derived a computation system to represent negative numbers with chemical species and reactions, the downside of this method is that it needs to double both the number of chemical species and reactions. We will therefore pursue an approximation of Eq.~\eqref{eq:Lhat} that does not give negative output. We propose to replace the RHS of Eq.~\eqref{eq:Lhat} by:
\begin{eqnarray}
g_- \; x_*(t) \times \left\{ \left[\log(\lambda_k(t)) - \frac{\lambda_k(t)}{u(t)} \right]_+ \right\} 
\label{eq:Ltilde_pre}
\end{eqnarray}
which is always non-negative. With reference to our example in Fig.~\ref{fig:local_1}, the consequence of this modification is that when Symbol 0 is used with Filter 1, the output is zero. This reduces the gap between the filter outputs for Symbol 0. Hence, the design decision is a trade-off between receiver complexity and demodulation accuracy. We remark that this strategy of using only positive log-probability can also be used to obtain molecular circuits to detect sustained signals \cite{Chou:2018jh}. 

Another difficulty of implementing the intermediate approximation is we need to find a way to encode the reference signal $\lambda_k(t)$ using chemical reactions. The difficulty is that $\lambda_k(t)$ is a dynamical signal. We now argue that Eq.~\eqref{eq:Ltilde_pre} is approximately equal to the RHS of: 
\begin{eqnarray}
\frac{d\tilde{L}_k(t)}{dt} = g_- \; x_*(t) \times \left\{ \left[\log(a_k) - \frac{a_k}{u(t)} \right]_+ \right\},  
\label{eq:Ltilde_pre2}
\end{eqnarray}
i.e. $\lambda_k(t)$ in Eq.~\eqref{eq:Ltilde_pre} is replaced by the constant $a_k$. First, we consider $t \in [0,d)$, the expression Eq.~\eqref{eq:Ltilde_pre} and the RHS of Eq.~\eqref{eq:Ltilde_pre2} are equal because $\lambda_k(t) = a_k$ for all $t$ in this time interval. For $t \geq d$, $\lambda_k(t) = u(t) = b$, therefore the term in curly brackets in Eq.~\eqref{eq:Ltilde_pre} equals $[\log(b)-1]_+$, which is small, and the term in curly brackets in Eq.~\eqref{eq:Ltilde_pre2} is zero because $a_k \gg b$. Therefore, Eq.~\eqref{eq:Ltilde_pre} and the RHS of Eq.~\eqref{eq:Ltilde_pre2} are approximately equal. The good news is that we can replace the dynamical reference signal by a constant, which is easier to realise by chemical reactions. Note that the RHS of Eq.~\eqref{eq:Ltilde_pre2} is zero for $t \geq d$, this means we are not using any information from $x_*(t)$ for $t \geq d$. We mentioned earlier that this is not an issue because $x_*(t)$ is uninformative in this duration. 
 
The last step is to approximate the factor in curly brackets in Eq.~\eqref{eq:Ltilde_pre2} by a Hill function \cite{Keener} to obtain filters of the form:
\begin{eqnarray}
\frac{dy_k(t)}{dt} = g_- \; x_*(t) \times \left\{ \frac{h_k u(t)^{n_k}}{H_k^{n_k} + u(t)^{n_k}} \right\}
\label{eqn:mol_rea}
\end{eqnarray}  
so that $y_k(t) \approx \tilde{L}_k(t)$ where $\tilde{L}_k(t)$ comes from Eq.~\eqref{eq:Ltilde_pre2}. The importance of Eq.~\eqref{eqn:mol_rea} is that it can be realised by chemical reactions because: Hill functions can be realised by chemical reactions\footnote{\color{black} \label{fn:hill1} Hill functions are often used in the mathematical biology literature to model reactions with cooperativity \cite{Keener}. For a given Hill function, there are many possible ways to realise it using chemical reactions. These realisations differ in terms of complexity, as well as dynamical and noise properties. In this paper, we will use Hill function as a ``block" and we will leave the problem of the selecting a good realisation as future work. Note that the analysis of chemical systems realising Hill function is challenging because of non-linearity.}, so we can view the factor in curly bracket as the concentration of a chemical species; and consequently the RHS of Eq.~\eqref{eqn:mol_rea} can be interpreted as the reaction rate of a bi-molecular reaction\footnote{\color{black} In order to understand this interpretation, let us consider the bi-molecular reaction \cee{A + B ->[k] C} where \cee {A}, \cee {B} and \cee{C} denote three chemical species, and $k$ is the reaction rate constant. This bi-molecular reaction can be modelled by the ODE $\frac{d[C](t)}{dt} = k \; [A](t) \;  [B](t)$ where $[A](t)$ is the concentration of the chemical species \cee{A} at time $t$ etc. By comparing this ODE against Eq.~\eqref{eqn:mol_rea}, we can identify $[C](t)$ with $y_k(t)$, $g_-$ with $k$. $[A](t)$ with $x_*(t)$ and $[B](t)$ with the Hill function. Hence the interpretation.}. In order to make $y_k(t) \approx \tilde{L}_k(t)$, we need to match the RHSs of Eq.~\eqref{eq:Ltilde_pre2} and Eq.~\eqref{eqn:mol_rea}. We propose to determine the Hill function parameters $h_k$, $H_{k}$ and $n_k$ such that the difference between the following two expressions is small in the least squares sense: 
\begin{eqnarray}
\left\{ \left[\log(a_k) - \frac{a_k}{q} \right]_+ \right\} \approx  \frac{h_k q^{n_k}}{H_k^{n_k} + q^{n_k}} 
\label{eq:hill_fit}
\end{eqnarray}
for $q > \frac{a_k}{\log(a_k)}$ and for $k = 0, ..., K-1$. We need to do this fitting $K$ times, one for each $a_k$. {\color{black} Note that for $q \leq \frac{a_k}{\log(a_k)}$, the left-hand side (LHS) of Eq.~\eqref{eq:hill_fit} is zero and for $q> \frac{a_k}{\log(a_k)}$, it is a strictly increasing function of $q$. We have therefore restricted the fitting to the range $q > \frac{a_k}{\log(a_k)}$ because both sides of Eq.~\eqref{eq:hill_fit} are strictly increasing functions of $q$ in this range. }

\begin{figure*}[t]
        \centering
        \begin{subfigure}[t]{0.45\textwidth}
        		\centering
        		\includegraphics[scale=0.40]{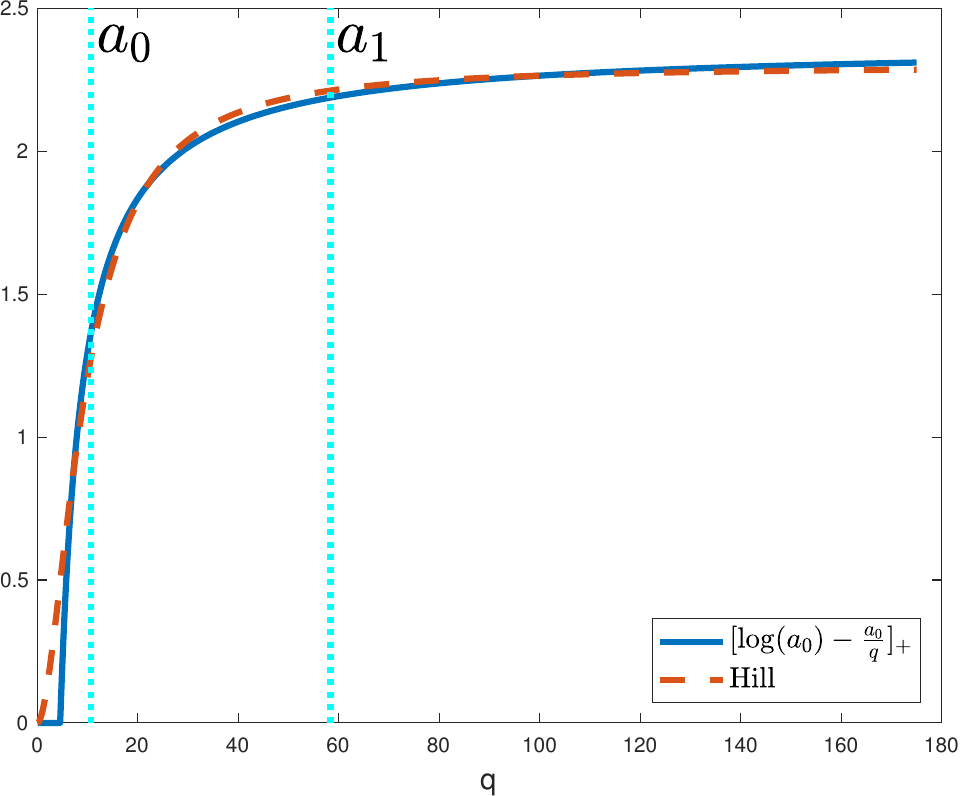}
        		\caption{Filter 0.}
        		\label{fig:hill_fit0}
	\end{subfigure} 	
	\begin{subfigure}[t]{0.45\textwidth}
        		\centering
        		\includegraphics[scale=0.40]{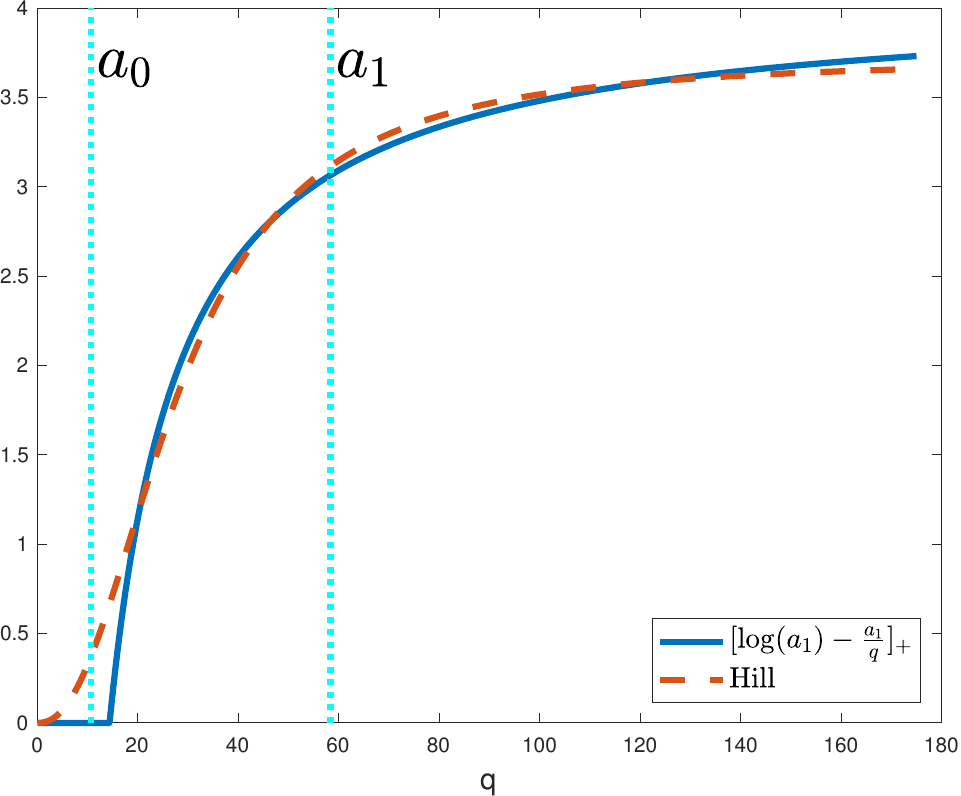}
        		\caption{Filter 1.}
        		\label{fig:hill_fit1}
	\end{subfigure} 	
\caption{Comparing the fit of Eq.~\eqref{eq:hill_fit} for Filters 0 and 1.}
\label{fig:hill_fit}	
\end{figure*}

\begin{figure*}[t]
    \centering
    \begin{subfigure}[t]{0.45\textwidth}
        \centering
        \includegraphics[scale=0.40]{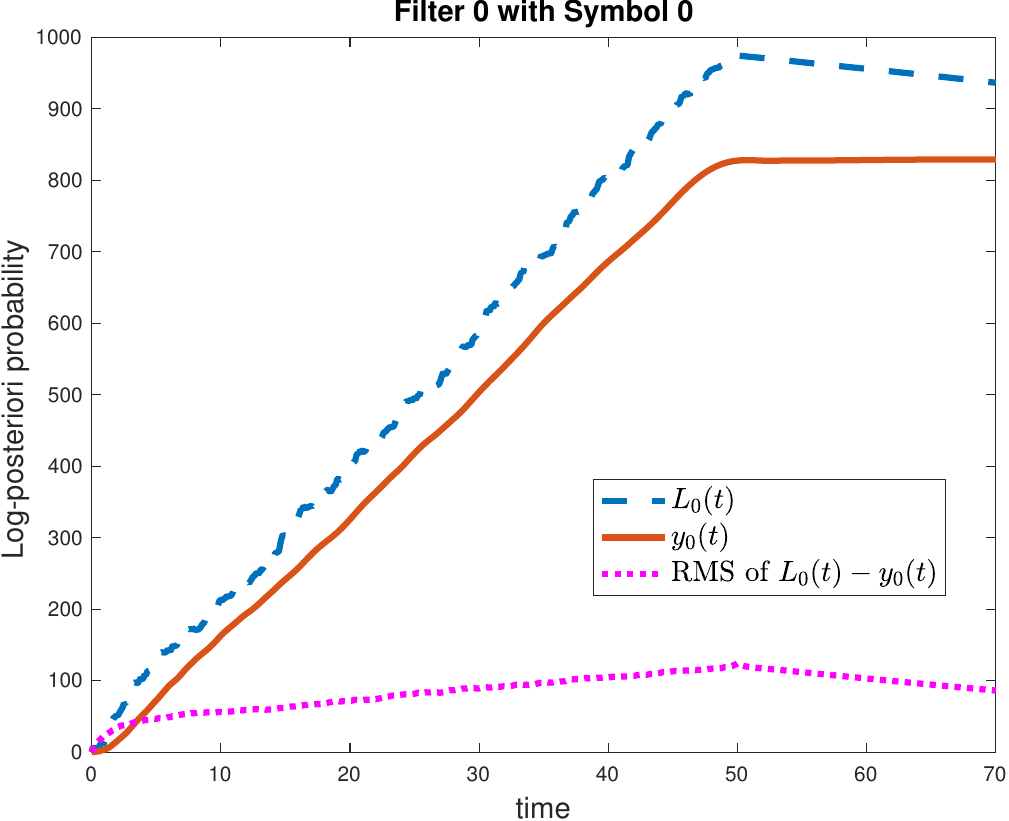}
        \caption{Filter 0 with Symbol 0.}
        \label{fig:local_2_00}
    \end{subfigure} 
    \begin{subfigure}[t]{0.45\textwidth}
        \centering
        \includegraphics[scale=0.40]{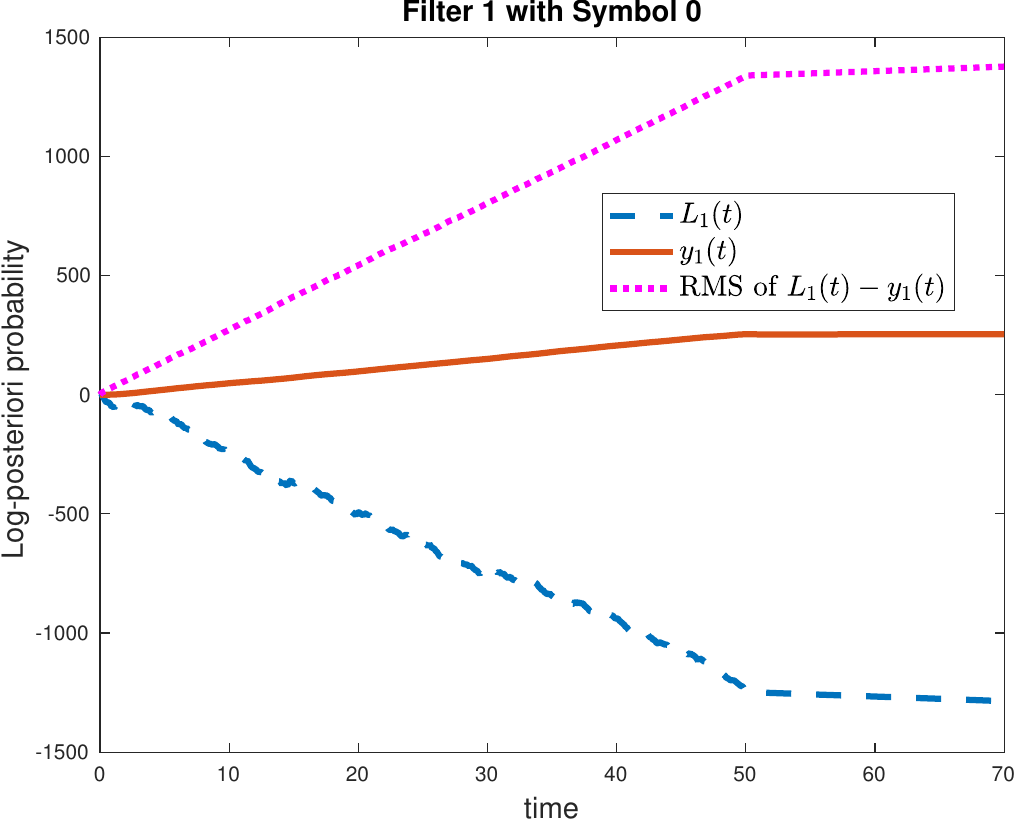}
        \caption{Filter 1 with Symbol 0.}
        \label{fig:local_2_01}
    \end{subfigure}   
   
    \begin{subfigure}[t]{0.45\textwidth}
        \centering
        \includegraphics[scale=0.40]{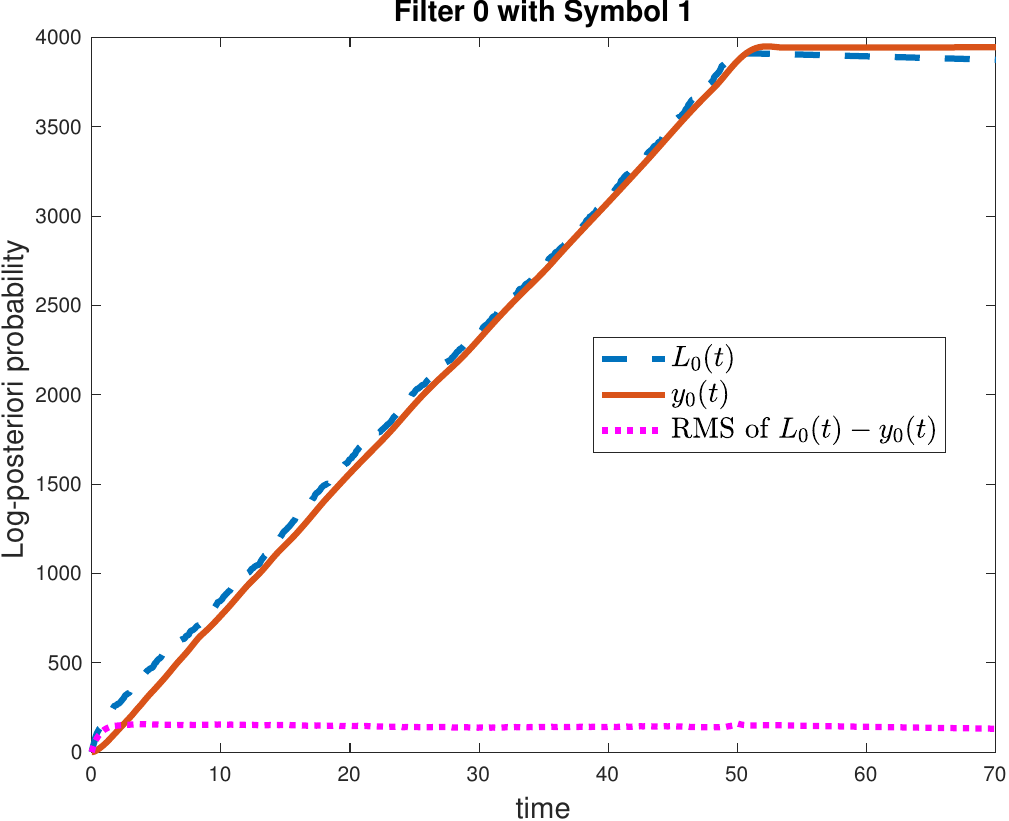}
        \caption{Filter 0 with Symbol 1.}
        \label{fig:local_2_10}
    \end{subfigure} 
    \begin{subfigure}[t]{0.45\textwidth}
        \centering
        \includegraphics[scale=0.40]{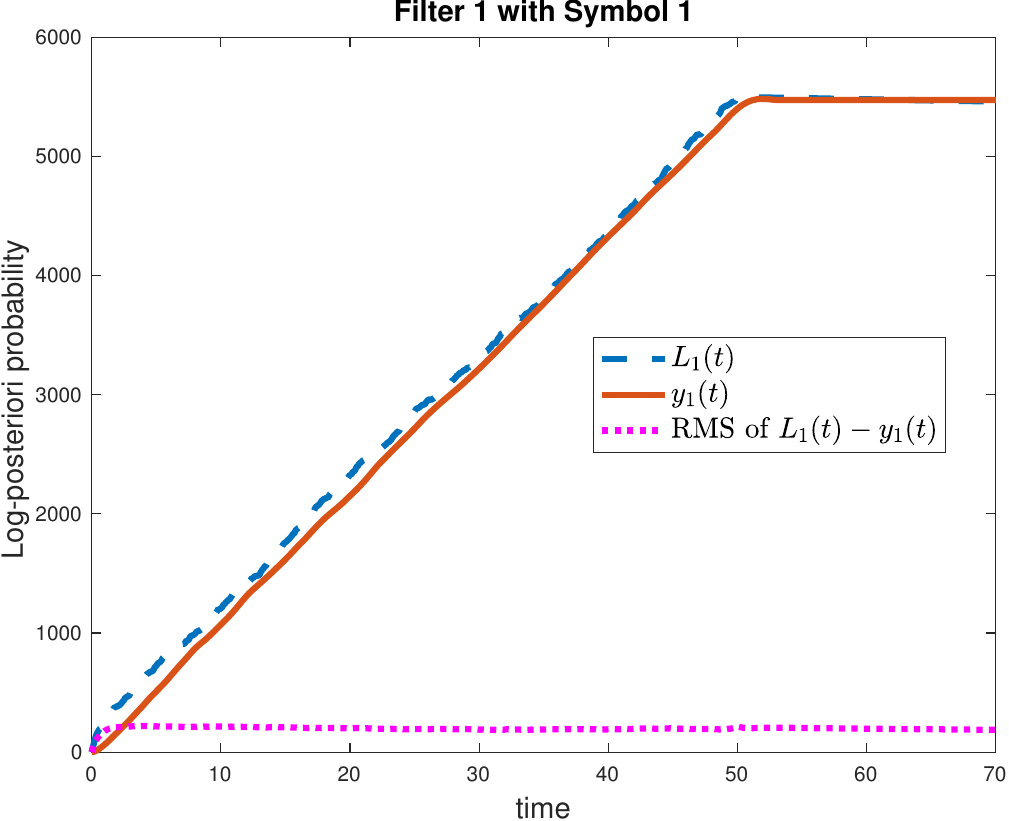}
        \caption{Filter 1 with Symbol 1.}
        \label{fig:local_2_11}
    \end{subfigure}    
\caption{Comparing $L_k(t)$, $y_k(t)$ and the RMS of $L_k(t)-y_k(t)$.}
\label{fig:local_2}
\end{figure*}

{\color{black}
We continue with the earlier numerical example. Fig.~\ref{fig:hill_fit} plots the two sides of Eq.~\eqref{eq:hill_fit} as a function of $q$ for Filters 0 and 1. It can be seen that the fit is good for sufficiently large $q$. Fig.~\ref{fig:local_2} compares the true log-posteriori probability $L_k(t)$ with the molecular approximation $y_k(t)$. The solid red lines show $y_k(t)$ from one realisation of $x_*(t)$ and the dashed blue lines show one realisation of $L_k(t)$. The dotted magenta lines show the RMS of $L_k(t)-y_k(t)$ computed from 100 independent simulations. It can be seen that if $L_k(t) \geq 0$, then $y_k(t)$ approximates $L_k(t)$; this applies to the following three cases: Filter 0 for both Symbols, as well as to Filter 1 for Symbol 1. These three cases also correspond to good fit of Eq.~\eqref{eq:hill_fit} in Fig.~\ref{fig:hill_fit}. Fig.~\ref{fig:hill_fit0} shows that for Filter 0, the fit is good at $a_0$ and $a_1$, which are respectively the amplitudes of Symbols 0 and 1. However for Filter 1, Fig.~\ref{fig:hill_fit1} shows only good fit at $a_1$, which corresponds to Symbol 1. Lastly, we consider the case of Symbol 0 with Filter 1. Fig.~\ref{fig:local_2_01} shows that $y_k(t)$ does not approximate $L_k(t)$ for this case. This is because $L_k(t)$ is negative and the approximation $y_k(t)$ is always non-negative. In order to better understand how this $y_k(t)$ is obtained, we trace through the approximations that we have made. The first approximation is Eq.~\eqref{eq:Ltilde_pre2} whose RHS, for this case, equals $\left[\log(a_1) - \frac{a_1}{a_0} \right]_+$, which is zero because $a_1 > a_0$. This is then followed by the fitting in Eq.~\eqref{eq:hill_fit} and we can see from Fig.~\ref{fig:hill_fit1} that the fitted Hill function has a small non-zero value at $a_0$. This explains why $y_k(t)$ has a small positive value for this case.} For each symbol, the demodulator should select the $k$ that maximises $y_k(t)$ as the estimation of the transmitter symbol. We can see from Fig.~\ref{fig:local_2} that we can correctly demodulate using $y_k(t)$. 

{\color{black} In this section, we make the assumption of long pulse duration in order to reduce the complexity of the chemical based demodulator. This assumption is used to derive the intermediate approximation \eqref{eq:Lhat} and to replace $\lambda_k(t)$ by a constant. Consequently, if this assumption is not made, then the demodulator has more dynamics which in turn means more chemical reactions are needed to realise the demodulator. This discussion shows that there is a tradeoff between the length of the symbol duration and the complexity of the receiver. The study of this tradeoff is an open research problem.}

{\color{black}
\begin{remark}
Note that the molecular realisation in Eq.~\eqref{eqn:mol_rea} is different from the one in our earlier conference publication \cite{Chou2018:nanocom}. The difference lies in the factor within the curly brackets $\{ \; \}$ on the RHS of Eq.~\eqref{eqn:mol_rea}. The factor in $\{ \; \}$ in Eq.~\eqref{eqn:mol_rea} is based on the signal $u(t)$ but the one in \cite{Chou2018:nanocom} is based on $x_*(t)$. This change is necessary so that we can relate Eq.~\eqref{eqn:mol_rea} to a real-life biochemical circuit in Sec.~\ref{sec:promotor}. 
\end{remark}
}

\begin{remark} 
Although the procedure presented in Secs.~\ref{sec:sep} and \ref{sec:mol0} works for CM, it does not work for Duration Modulation (DM). In DM, Symbol $k$ is a rectangular pulse of duration $d_k$, with an amplitude $\tilde{a}$ when it is ON and $b$ when it is OFF. Let us consider DM with two symbols where Symbol 0's duration is shorter than Symbol 1's. It can be shown that: (i) the intermediate approximation Eq.~\eqref{eq:Lhat} is poor when Filter 1 is used with Symbol 0; (ii) the approximation Eq.~\eqref{eq:Ltilde_pre2} does not hold when Filter 0 is used with Symbol 1. This shows the limitation of the approximation procedure presented earlier. However we can show that the procedure can be modified to derive a chemical reaction-based demodulator for DM with two symbols. The two key modifications needed are: (i) Replacing log posteriori probability by log posteriori probability ratio; (ii) Instead of encoding the reference signals $\lambda_k(t)$ using constants, we need to encode a dynamical signal. The details can be found in \cite{Chou:2018jh}. 
\end{remark} 

\section{A biochemical circuit that behaves as Eq.~\eqref{eqn:mol_rea}}
\label{sec:promotor} 
\ifarxiv 
We show in Sec.~\ref{sec:mol0} that $y_k(t)$ in Eq.~\eqref{eqn:mol_rea} can be interpreted as approximate positive log posteriori probability. We also argue that it is possible to realise Eq.~\eqref{eqn:mol_rea} by using chemical reactions. However, in order to implement Eq.~\eqref{eqn:mol_rea}, we still need to point out the chemical species and chemical reactions that are needed. This aim of this section is to address this gap. 

Eq.~\eqref{eqn:mol_rea} is an ODE or an analog filter. There is much recent work in molecular computing and synthetic biology on realising analog computation. We can divide the work into two categories. The first category, which is based on a class of chemical reaction known as the strand displacement reaction (SDR) \cite{Soloveichik:2010be}, aims to provide a generic method of implementing analog computation. For example, \cite{Oishi:2011ig} presented a systematic method to implement any ODE with constant coefficients using SDR. In addition, SDR can also be used to realise arithmetic operators and rational functions, see e.g. \cite{Buisman:ub,Salehi:2016dw,Chou:2017bx}. The key advantage of SDR is that it is a general methodology which works for many types of analog computation. The downside is that SDR often requires many more chemical reactions than what is minimally required, e.g. for a rational function that approximately computes logarithm, Table 1 in \cite{Chou:2017bx} shows that theoretically it is possible to realise the calculation with only 7 species and 13 reactions, but an SDR implementation requires 62 species and 40 reactions. 

Note that SDR is based on chemicals that are artificially synthesised, i.e. not commonly found in living cells. In contrast, the second category of work is to use chemical species that are naturally found in living cells to do analog computation. For example, the authors in \cite{Daniel:2013ke} derived a number of molecular circuits that can compute logarithm using proteins and DNA (deoxyribonucleic acid) that are found inside the bacteria Escherichia coli. The advantage of this category of work is that the circuits they produce often require few chemical species. However, systematic design methods do not appear to exist at this moment.  

Given the flexibility of SDR, it is possible to implement Eq.~\eqref{eqn:mol_rea} using SDR and we therefore will not consider it here. In this section, we want to explore the possibility of using chemical species found in living cells to implement Eq.~\eqref{eqn:mol_rea}. In this section, we will show that a gene promotor circuit named DCS2, which is found in Saccharomyces cerevisiae (yeast) and was studied in \cite{Hansen:2013fs}, behaves similarly to the CM demodulation filter Eq.~\eqref{eqn:mol_rea}. Note that in order to implement the modulator in Fig.~\ref{fig:demod}, we need as many demodulation filters as the number of symbols. Since DCS2 can only be used to implement one demodulation filter, we do not have a complete design. The purpose of this section is therefore not to present a complete design, but rather to present a pointer to show where one may try to search for the chemical species and chemical reactions that can implement the demodulation filters.  

We will first provide some background on DCS2 in Sec.~\ref{sec:dcs2_bg}. We then, in Secs.~\ref{sec:dcs2_fitting} and \ref{sec:dcs2_evi}, use the data from \cite{Hansen:2013fs} to demonstrate that DCS2 and Eq.~\eqref{eqn:mol_rea} have similar behaviour. 

\fi

\subsection{Background on DCS2} 
\label{sec:dcs2_bg} 
We begin by presenting some background on DCS2 and the experiments performed in \cite{Hansen:2013fs}. We will be using some terminology in molecular biology in this description. We are conscious that some of the readers of this publication are not familiar with these terminologies but at the same time we want to cater for those who are conversant in them. In the following, the molecular biology terminologies will be typeset in italics and reader who are unfamiliar with them can simply think of them as the name for a class of biochemical molecules. 

The {\sl gene promotor} DCS2 can react with the {\sl transcription factor} Msn2 to turn DCS2 into an active state. The authors of \cite{Hansen:2013fs} derived an experimental technique to indirectly manipulate the concentration of Msn2 over time using an {\sl inhibitor molecule} 1-NM-PP1. Their aim was to understand how DCS2 behaved with different time-varying concentration profiles of Msn2. In other words, one can view the time profile of Msn2 as a time-varying input. 

The authors of \cite{Hansen:2013fs} used 30 different time profiles of Msn2 as the input. Time profiles 1--20 were rectangular pulses of different amplitudes and durations; more specifically, they came from the Cartesian product of a set 1-NM-PP1 amplitudes \{100nM,275nM,690nM,3$\mu$M\} and a set of durations \{10,20,30,40,50\} in minutes. Note that we can interpret these 20 time profiles as 5 sets of CM experiments, where each experiment consisted of a pulse duration and 4 amplitudes. Time profiles 21--26 consisted of different number of 1-NM-PP1 pulses with amplitudes 690nM; the number of pulses used were 2, 3, 4, 5, 6 and 8. Lastly, time profiles 27--30 consisted four 1-NN-PP1 pulses at 690nM but different time spacing between the pulses. 

When DCS2 is active, it will enable some other chemical reactions to occur. One of these reactions will produce {\sl messenger ribonucleic acid} mRNA. The authors of \cite{Hansen:2013fs} designed this mRNA so that it will produce {\sl yellow fluorescent protein} YFP. After YFP has matured, it becomes a {\sl matured yellow fluorescent protein} mYFP. The experiments measured the intensity of mYFP which indicates the concentration of mRNA.  One may therefore consider the time series of Msn2 and mYFP as, respectively, the input and output of a system of chemical reactions.

{\color{black} For each of the Msn2 time profile, appropriate quantity of the {\sl inhibitor molecule} 1-NM-PP1 was administered to a number of yeast cells in order to realise the specific Msn2 time profile in these cells. For each cell, a time series of its mYFP intensity is recorded with a sampling interval of 2.5 minutes. Each time series consists of 64 data points.  For each of the Msn2 time profile, all the mYFP time series obtained are averaged to obtained one time series. The authors of \cite{Hansen:2013fs} used the averaged time series for model fitting. 
}

The sequence of chemical reactions from Msn2 to the production of mRNA is unknown while those from mRNA to mYFP are known. In other words, the dynamical model from Msn2 to mRNA is unknown. In order to determine this unknown model, the authors of \cite{Hansen:2013fs} used a number of different models and checked which one of them gave the best fit to the Msn2 and mYFP data series. Note that none of the models used in \cite{Hansen:2013fs} had a connection with the CM model in Sec.~\ref{sec:mol0}. 

\subsection{Model fitting}
\label{sec:dcs2_fitting} 

\begin{figure*}[t]
    \centering
    \begin{subfigure}[t]{0.45\textwidth}
        \centering
        \includegraphics[scale=0.40]{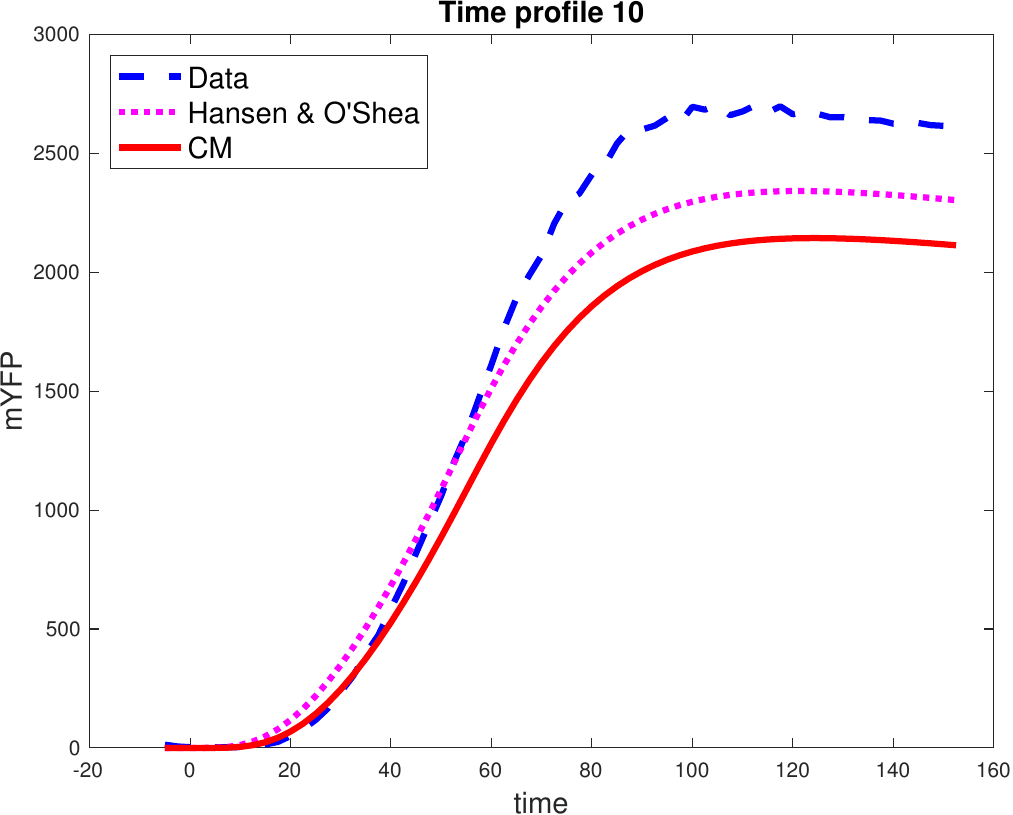}
        \caption{Input: Rectangular pulse with amplitude 275nM, duration 50 minutes.} 
        \label{fig:bio_1_1}
    \end{subfigure} 
    \begin{subfigure}[t]{0.45\textwidth}
        \centering
        \includegraphics[scale=0.40]{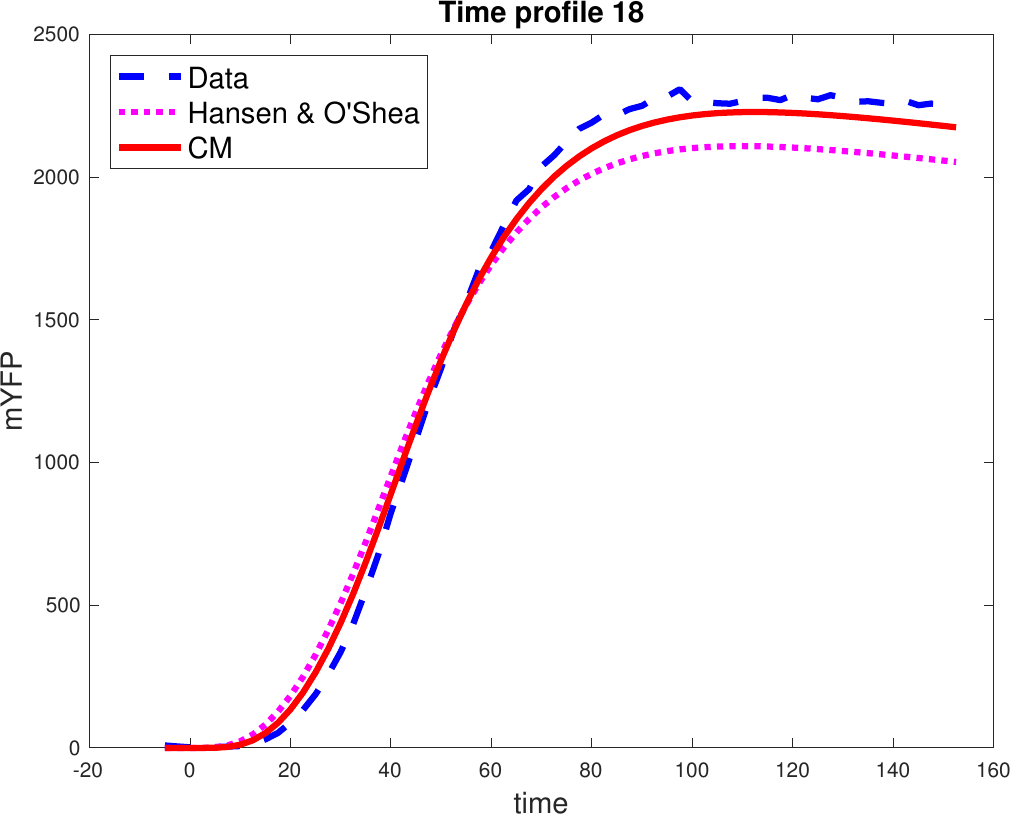}
        \caption{Input: Rectangular pulse with amplitude 3$\mu$M, duration 30 minutes. } 
        \label{fig:bio_1_2}
    \end{subfigure}   
   
    \begin{subfigure}[t]{0.45\textwidth}
        \centering
        \includegraphics[scale=0.40]{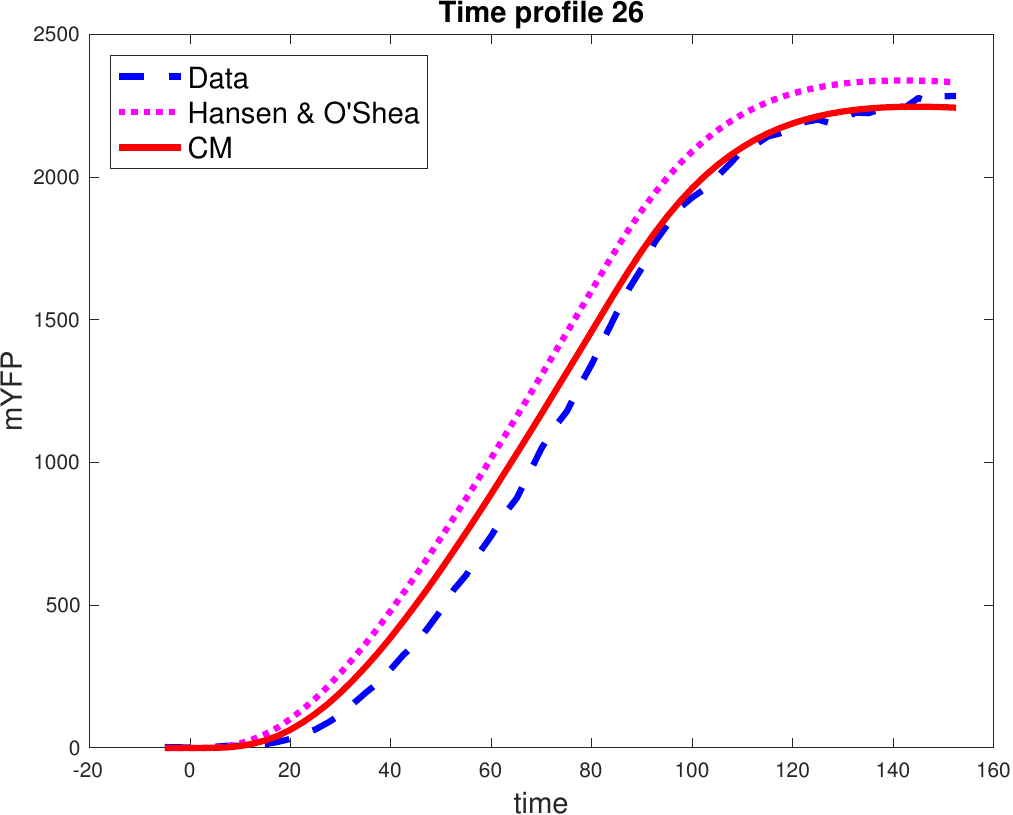}
        \caption{Input: 6 pulses of amplitude 690nM and duration 5 minutes.} 
        \label{fig:bio_1_3}
    \end{subfigure} 
    \begin{subfigure}[t]{0.45\textwidth}
        \centering
        \includegraphics[scale=0.40]{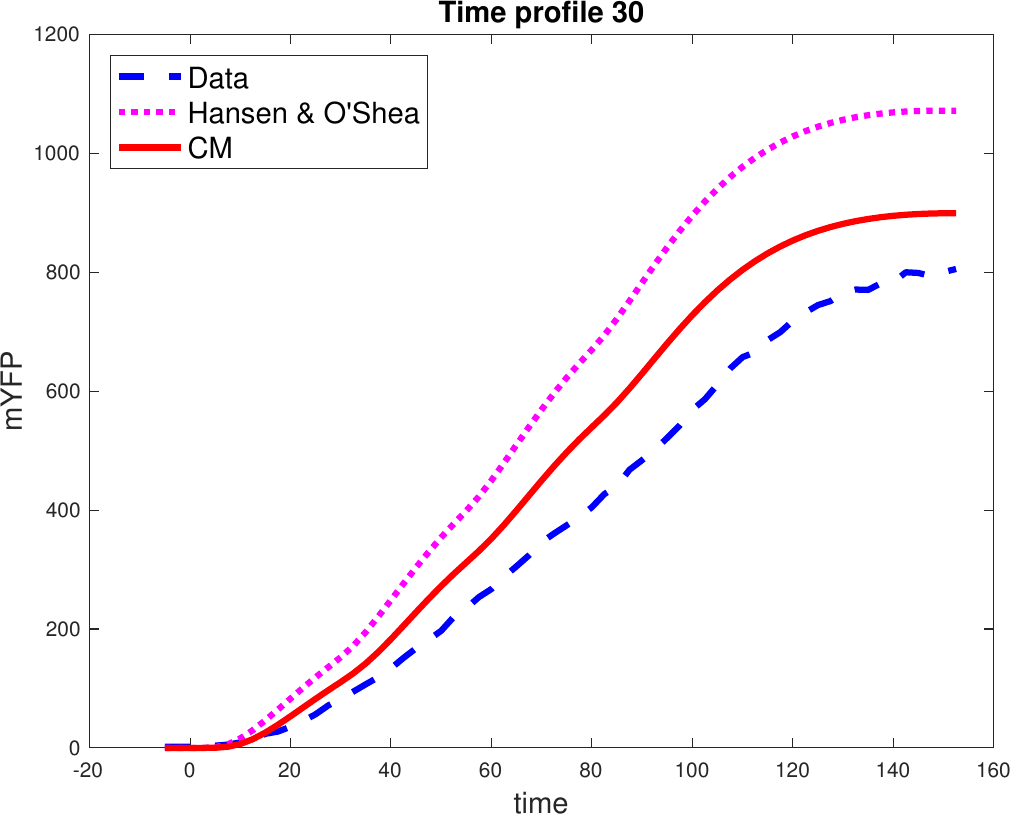}
        \caption{Input: 4 pulses of amplitude 690nM and duration 5 minutes.} 
        \label{fig:bio_1_4}
    \end{subfigure}    
\caption{This figure compares the measured mYFP against the predicted mYFP from two models. The dashed blue lines show the measured data. The magenta dotted lines show the predicted mYFP from the model from \cite{Hansen:2013fs}. The red solid lines show the predicted mYFP for our CM-inspired model. Subfigure (a) is an example that our model gives a worse fit. Our model gives a better fit for the other 3 subfigures. Fitting errors for the model from \cite{Hansen:2013fs} for the four subfigures ($\times 10^6$): 4.02, 1.34, 1.87, 3.31.  Fitting errors for our model for the four subfigures ($\times 10^6$): 10.8, 0.24, 0.39, 0.67.}
\label{fig:bio_1}
\end{figure*}


Since the experiments conducted in \cite{Hansen:2013fs} were based on CM, we are curious whether DCS2 may behave like a CM demodulation filter. In this section, we will fit a CM-inspired model to the Msn2-mYFP experimental data\footnote{The data are available at \url{https://anderssejrhansen.wordpress.com/data/}.}. In our model, we identify Msn2 and DCS2 with, respectively, ${\cee S}$ and ${\cee X}$ in Eq.~\eqref{cr:all}; and they react according to chemical reactions \eqref{cr:all}. We will refer to the counterpart of ${\cee X_*}$ as active DCS2. Let $P_{\rm active}(t)$ denote the fraction of DCS2 that is active at time $t$. The reactions in \eqref{cr:all} can be modelled by:
\begin{eqnarray}
\frac{d P_{\rm active}(t)}{dt} =  g_+ [Msn2](t) \; (1 - P_{\rm active}(t)) - g_-  \; P_{\rm active}(t) \label{eq:cm_bio_1} 
\end{eqnarray} 
where $[Msn2](t)$ denotes the concentration of Msn2 at time $t$. Note that the analysis in \cite{Hansen:2013fs} suggested that the reactions between DCS2 and Msn2 are fast. 

Inspired by the CM demodulation filter Eq.~\eqref{eqn:mol_rea}, we propose the following equation to model the reaction that active DCS2 triggers: 
\begin{eqnarray}
\frac{d [C_{\rm init}](t)}{dt} & = & g_- \; P_{\rm active}(t) \times \left\{ \frac{h [Msn2](t)^{n}}{H^{n} + [Msn2](t)^{n}} \right\} \nonumber \\  & & - d_2 [C_{\rm init}](t)
\label{eq:cm_bio_2} 
\end{eqnarray}  
where $h$, $n$ and $H$ are coefficients of Hill function; we remark that we have for simplicity dropped the subscript $k$ because we are considering only one demodulation filter. Note that the first term on the RHS of the above equation is similar to the RHS of Eq.~\eqref{eqn:mol_rea} because we identify $P_{\rm active}(t)$ and $[Msn2](t)$ with, respectively, $x_*(t)$ and $u(t)$. Since $P_{\rm active}(t)$ is an active {\sl gene promotor}, this term can be viewed as the reaction rate of a type of reaction called the transcription initiation process, see \cite{Hao:2011kz}. So, in the above equation, $[C_{\rm init}](t)$ is the concentration of the {\sl initiation complex} and $d_2$ is its degradation rate constant. 

Recall that the Hill function coefficients in Eq.~\eqref{eqn:mol_rea} are obtained from performing a least-squares fit, see Eq.~\eqref{eq:hill_fit}. This means the Hill function coefficients in Eq.~\eqref{eqn:mol_rea} are {\bf dependent} on an amplitude parameter $a_k$. Therefore, for our CM-inspired model, we assume the Hill function coefficients $h$, $n$ and $H$ are related to an amplitude parameter $a$ via a least-squares fit: 
\begin{eqnarray}
\left\{ \left[\log(a) - \frac{a}{q} \right]_+ \right\} \approx  \frac{h q^{n}}{H^{n} + q^{n}} 
\label{eq:cm_bio_2_hill}
\end{eqnarray}
and the fit should hold for $q > \frac{a}{\log(a)}$. 

The next reaction is the production of mRNA from  the {\sl initiation complex}. According to \cite{Hao:2011kz}, this can be modelled by:
\begin{eqnarray}
\frac{d [mRNA](t)}{dt} & = & k_3 [C_{\rm init}](t) - d_3 [mRNA](t) \label{eq:cm_bio_3} 
\end{eqnarray} 
where $[mRNA](t)$ is the concentration of mRNA, and $k_3$ and $d_3$ are reaction rate constants. 

Eqs.~\eqref{eq:cm_bio_1}--\eqref{eq:cm_bio_3} form our model from Msn2 to mRNA; for the model from mRNA to mYFP, we use Eqs.~(5)-(6) in \cite{Hansen:2013fs} 
which are in Eqs.~\eqref{eq:app:cm_bio_4}-\eqref{eq:app:cm_bio_5} in Appendix \ref{app:mYFP}. All these 6 equations together form our CM-inspired model. The unknown parameters for our model are $g_+$, $g_-$, $a$, $d_2$ and $k_3$; these 5 parameters will be used for fitting. Note that the Hill function coefficients $h$, $n$ and $H$ are {\bf not} free parameters and they implicitly depend on the {\bf free} parameter $a$. In other words, one can view Eq.~\eqref{eq:cm_bio_2_hill} as an ``equality" constraint in the model. Finally, we should point out that $d_3$ in Eq.~\eqref{eq:cm_bio_3} is also not a free parameter. This is consistent with \cite{Hansen:2013fs} where $d_3$ was obtained from an independent experiment rather than via model fitting. 

We fit our CM-inspired model to the Msn2 and mYFP data from \cite{Hansen:2013fs}. The authors in \cite{Hansen:2013fs} used all 30 sets of time profiles for fitting so we do the same for a fair comparison. The values of the optimised parameters for our model are: $g_+ = 3.19\times10^{-4}$, $g_- = 0.15$, $a = 1400$, $d_2 = 0.40$ and $k_3 = 0.23$. The fitting error is $3.9 \times 10^7$. The best model obtained by \cite{Hansen:2013fs} is given in Equations (1)-(6) in \cite{Hansen:2013fs}. Their model has \st{8} 7 parameters and a fitting error of $4.9 \times 10^7$ \cite[Supplementary Table 2]{Hansen:2013fs}. This shows that our CM-inspired model gives a better fit with fewer parameters. Out of the 30 datasets, our CM-inspired model is able to improve the fit for 23 of them. For illustration, Fig.~\ref{fig:bio_1} shows the mYFP for 4 input signals, where the dashed blue lines show the measured mYFP, the dotted magenta lines show the output of the model from \cite{Hansen:2013fs} and the red solid lines show the output from our model. Other than subfigure (a), our model gives a better fit. This numerical experiment therefore provides some evidence to show that DCS2 behaves like the demodulation filter Eq.~\eqref{eqn:mol_rea}. {\color{black} To the best of our knowledge, the biochemistry of how DCS2 decodes (or how cells decode) CM signals is still an open problem. Our proposed model may help to answer this question.} 

\subsection{Further evidence that DCS2 behaves like Eq.~\eqref{eqn:mol_rea}}
\label{sec:dcs2_evi}
Note that all the Msn2 input time profiles in \cite{Hansen:2013fs} consists of one or more rectangular pulses with two amplitude levels, a high ON-amplitude and a zero OFF-amplitude. Under these types of input and the assumption that the reactions between DCS2 and Msn2 are fast, we show in Appendix \ref{app:mYFP} that, if such signals are used as the input to our CM-inspired model for DCS2, then for a given ON-amplitude, the maximum mYFP predicted by the model has the property that it is proportional to the total duration that the input signal is ON. For a rectangular pulse, the total duration that the input signal is ON is simply the duration of the pulse. Recall that the Msn2 signal for Time Profiles 21--30 consist of a train of 5-minute pulses (produced by an 1-NM-PP1 amplitudes of 690nM), then the total duration that the input signal is ON is the number of pulses times 5 minutes. We will first demonstrate that the data supports this proportionality property. 

\begin{figure}[t]
        \centering
    \begin{subfigure}[t]{0.45\textwidth}
        \centering
        \includegraphics[scale=0.40]{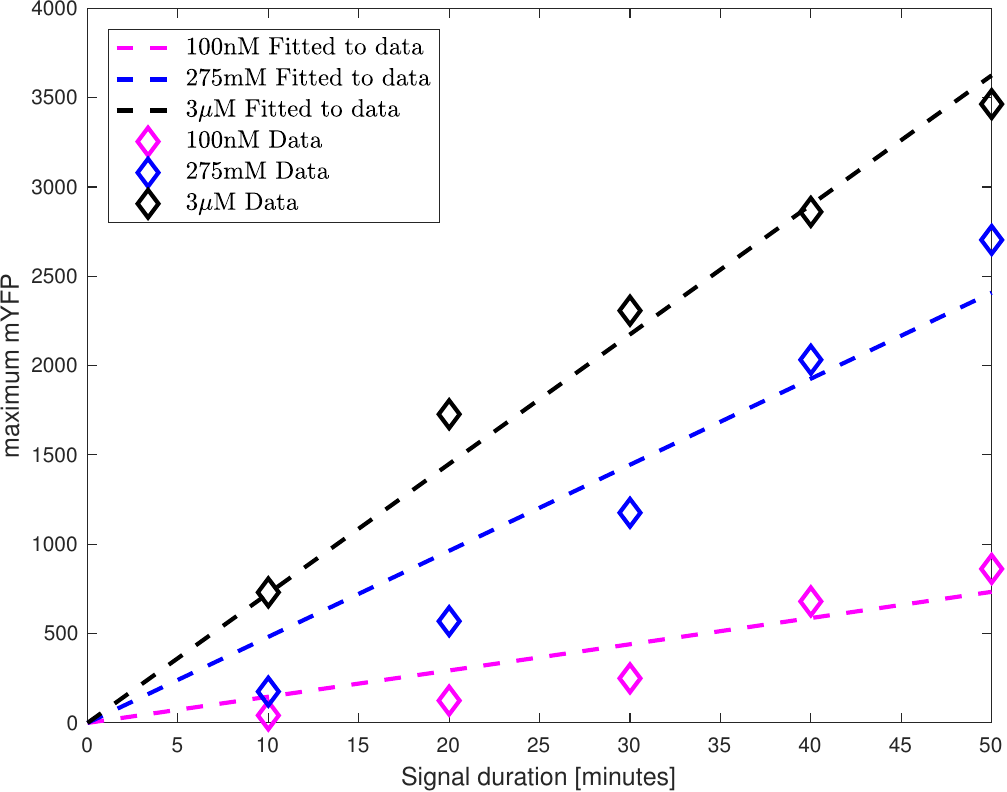}
        \caption{Maximum mYFP for varying Msn2 pulse durations for amplitudes 100nM, 275nM, 690nM and 3$\mu$M.} %
        \label{fig:bio_2_1}
    \end{subfigure} 
    \begin{subfigure}[t]{0.45\textwidth}
        \centering
        \includegraphics[scale=0.40]{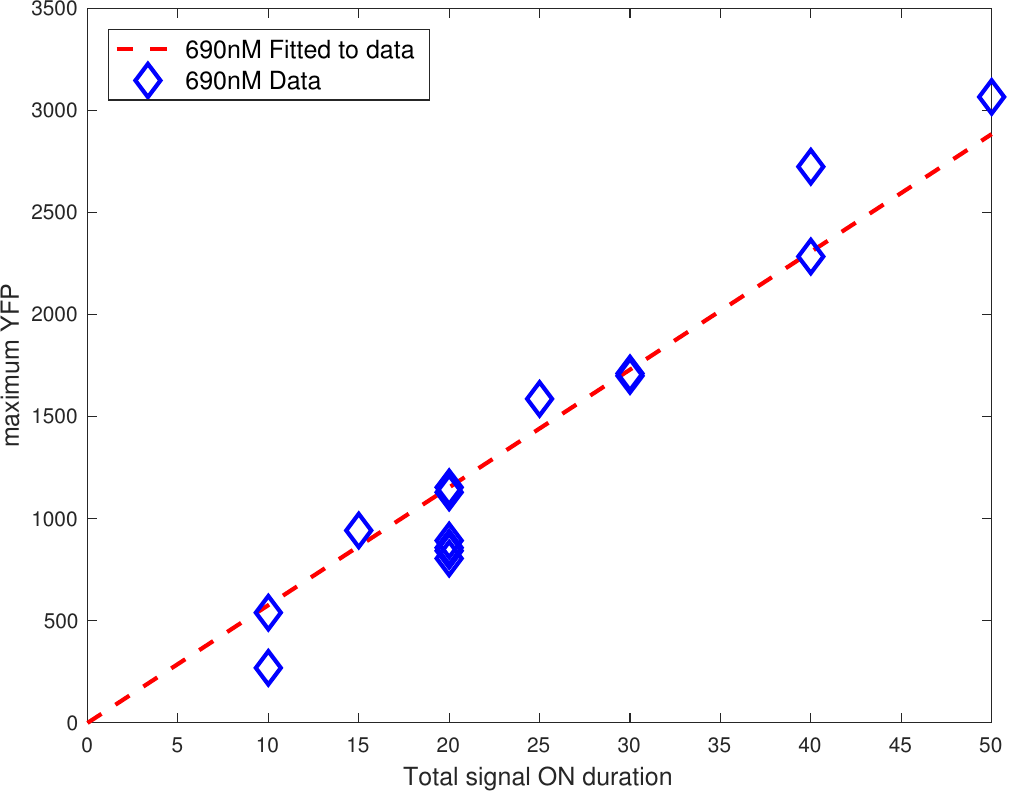}
        \caption{Maximum mYFP for varying Msn2 ON durations for amplitude 690nM.} %
        \label{fig:bio_2_2}
    \end{subfigure}            
    \caption{Demonstrating that the maximum mYFP is proportional to the duration that the Msn2 input is ON.}
    \label{fig:bio_2}
\end{figure}

The 30 datasets in \cite{Hansen:2013fs} are based on 4 different 1-NM-PP1 amplitudes of 100nM, 275nM, 690nM and 3$\mu$M which correspond to Msn2 amplitudes of respectively, 313.2, 744.5, 1107.8 and 1410.1. We first consider the amplitudes of 100nM, 275nM and 3$\mu$M. where the time profiles corresponding to these amplitudes have one rectangular pulse. For each amplitude, we plot the maximum mYFP against the duration of the pulse in Fig.~\ref{fig:bio_2_1} where the measured data are plotted using diamond markers. In order to demonstrate that, for a given amplitude, the mYFP is proportional to the pulse duration, we fit a straight line through the origin, see the dashed lines in Fig.~\ref{fig:bio_2_1}. It can be seen that mYFP data do lie approximately on a line for a given input amplitude. 

For the case of 690nM amplitude, there are time profiles consisting of only one pulse as well as multiple pulses. In Fig.~\ref{fig:bio_2_2}, we plot the measured mYFP against the total ON-duration of the input, see the diamond markers. We fit a straight line through the origin, see the dashed line. Again, the property seems to hold. 

\subsection{Fitting of other promoters}
We have also fitted our model to the other six promoters in \cite{Hansen:2013fs}. Table \ref{tab:fit_promoter} compares the fit of the model in \cite{Hansen:2013fs} against our proposed model. We have reported the fit to 1 decimal point and have used the red colour to indicate the model that has a lower fit. It can be seen that our model has better fit for 4 out of 6 promoters (SIP18, TKL2, RTN2, DDR2), and equally good fit for HKX1. Although the model in \cite{Hansen:2013fs} has a better fit for ALD3 but our model is only marginally worse. 

\begin{table} 
	\centering
	\begin{tabular}{| l | r | r |} \hline 
	                & \multicolumn{2}{c|}{Model fit}   \\ \hline
	Promoter & \cite{Hansen:2013fs} & Our model  \\ \hline
    SIP18  &  $2.3 \times 10^7$   & ${\color{red}{2.1\times 10^7}}$   \\ \cline{1-3}    
    ALD3  &   ${\color{red}{2.3\times 10^7}}$  &   $2.4 \times 10^7$  \\ \cline{1-3}
    TKL2   & $1.3 \times 10^7$   & ${\color{red}{0.9\times 10^7}}$   \\ \hline
    RTN2  &   $6.6 \times 10^7$  &   ${\color{red}{4.4\times 10^7}}$    \\ \hline
    DDR2  &   $9.0 \times 10^8$   &  ${\color{red}{4.7\times 10^8}}$   \\ \cline{1-3}
    HXK1  &   $2.4 \times 10^8$  &  $ 2.4 \times 10^8$     \\ \hline
	\end{tabular}
	\caption{Comparing the model fit of the promoters SIP18, ALD3, TKL2, RTN2, DDR2, HXK1.} 
	\label{tab:fit_promoter} 
\end{table}

\section{Molecular demodulator for diffusion-based molecular communication} 
\label{sec:full} 

\begin{figure}[t]
        \centering
        \includegraphics[scale=0.40]{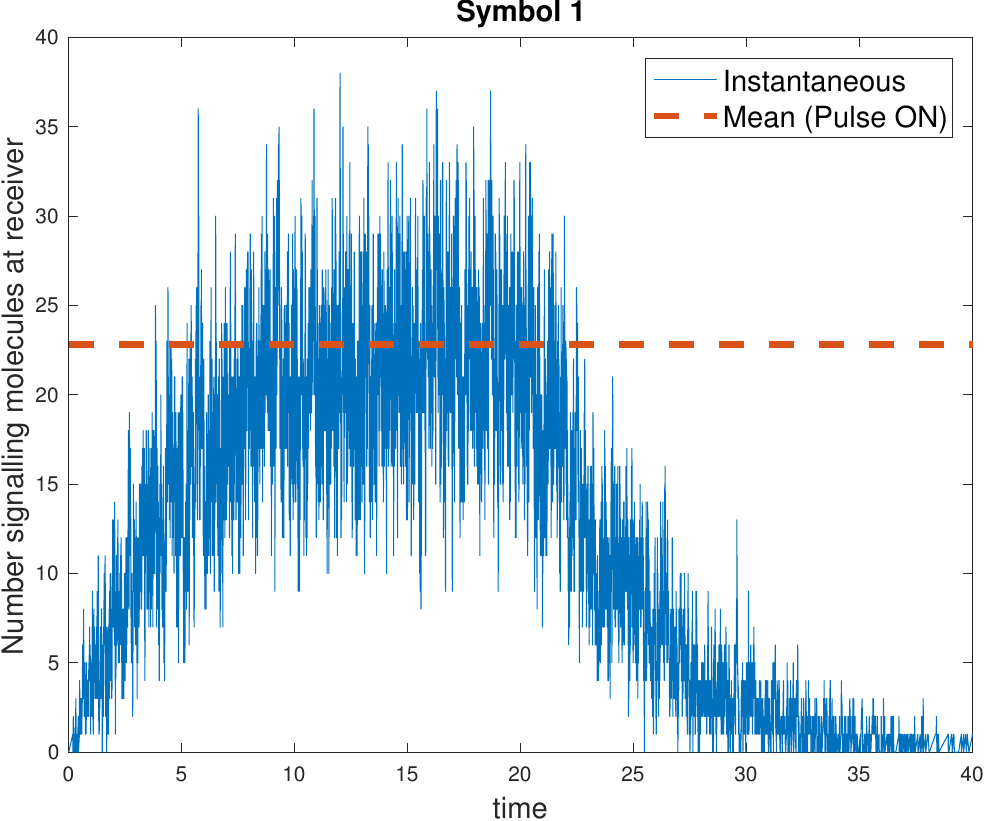}
       \caption{Instantaneous number of signalling molecules in the receiver voxel. The red line shows the mean number of signalling molecules in the receiver voxel assuming that the transmitter sends molecules continuously.}
        \label{fig:z_pulse}
\end{figure}

The molecular demodulation filter derived in Sec.~\ref{sec:simp} assumes that the transmitter and receiver are co-located, i.e. diffusion of signalling molecules were not considered. In this section, we will adapt the results in Sec.~\ref{sec:simp} for diffusion-based molecular communication. 

We assume the transmitter uses CM. Symbol $k$ emits $r_k$ number of signalling molecules per unit time when it is ON and at a basal rate when it is OFF. Due to diffusion, the number of signalling molecules at the receiver fluctuates and is no longer a rectangular pulse. For example, Fig.~\ref{fig:z_pulse} shows the number of signalling molecules at the receiver voxel. Although the fluctuation is clearly visible, we see that it is still possible to approximate the low frequency component of the signal by a rectangular pulse. This can be achieved by choosing the pulse duration to be sufficiently long. Therefore we can use this rectangular pulse approximation and the method in Sec.~\ref{sec:simp} to derive the demodulation filters. We see in Sec.~\ref{sec:mol0} that the derivation of the molecular demodulation filter requires only one parameter of the pulse: the amplitude of the pulse when it is ON, which is denoted as $a_k$ in Sec.~\ref{sec:simp}. We propose that, for Symbol $k$, we derive the molecular demodulation filter by assuming that $a_k$ is the mean number of molecules in the receiver voxel when the transmitter emits molecules at a rate of $r_k$ continuously.  For example, in Fig.~\ref{fig:z_pulse}, the thick solid red horizontal line shows the mean number of molecules in the receiver voxel if the transmitter emits continuously. 

We have now derived the molecular demodulation filters in Fig.~\ref{fig:demod}. The only component that we have not addressed is the maximum function in the demodulator in Fig.~\ref{fig:demod}. The computation of maximum is iterative and will require many additional chemical species and reactions. We propose an alternative which requires no new species and only one additional chemical reaction if $K = 2$. Let $\cee{Y_0}$ and $\cee{Y_1}$ be the chemical species which are represented by the output of the molecular demodulation filters $y_0(t)$ and $y_1(t)$. If we incorporate the chemical reaction 
\begin{align}
\cee{
Y_0 + Y_1 ->[k_a] \phi
}
\label{cr:anni} 
\end{align} 
which means the molecules $\cee{Y_0}$ and $\cee{Y_1}$ annihilate each other and if this reaction is fast, then only the species with a higher molecular count will remain. Since the molecular count is related to log-posteriori probability, this means that only the species corresponding to larger log-posteriori probability remains. Therefore, we can deduce the transmitted symbol by observing the type of chemical species that remains. 

Although the above method works for $K = 2$, the generalisation to $K \geq 3$ is problematic. 
\ifarxiv
See Appendix \ref{app:anni} for a more detailed discussion. We will leave the $K \geq 3$ case as future work. 
\fi 

\subsection{Numerical illustration}
We consider a medium of 2$\mu$m $\times$ 2$\mu$m $\times$ 1 $\mu$m. We assume a voxel size of $W^3$ $\mu$m$^{3}$  where $W = \frac{1}{3}$, creating an array of $6 \times 6 \times 3$ voxels. The transmitter and receiver are located at (0.5,0.8,0.5) and (1.5,0.8,0.5) (in $\mu$m) in the medium. The voxel co-ordinates are (2,3,2) and (5,3,2) respectively. We assume the diffusion coefficient $D$ of the medium is 1 $\mu$m$^2$s$^{-1}$. We assume an absorbing boundary for the medium and the signalling molecules escape from a boundary voxel surface at a rate of $\frac{D}{50W^2}$.

The propensity parameters of the receptors are: $g_+ = \frac{0.005}{W^3}$ s$^{-1}$ and $g_- = 1$ s$^{-1}$. {\color{black} The values of $D$, $g_+$ and $g_-$ are the same as those used in \cite{Chou:2015ga}. These values are similar to those used in \cite{Erban:2009us} and are considered to be biologically realistic.} 

We use $K = 2$ symbols. Each symbol is ON for 20s. When the symbol is ON, Symbol 0 (resp. Symbol 1) generates 150 (600) molecules per second. We assume Symbols 0 and 1 are sent with equal probability. 

We first consider the case of $M = 40$ receptors. We use the SSA algorithm to generate the time series of the number of active receptors up to 40s. Two time series are generated, one for Symbol 0 and the other for Symbol 1. We determine the mean number of signalling molecules in the receiver voxel assuming that the transmitter does not turn OFF and use this value as the $a_k$ for designing the molecular demodulation filter in Eq.~\eqref{eqn:mol_rea}. 

\begin{figure*}[t]
    \centering
    \begin{subfigure}[t]{0.45\textwidth}
        \centering
        \includegraphics[scale=0.40]{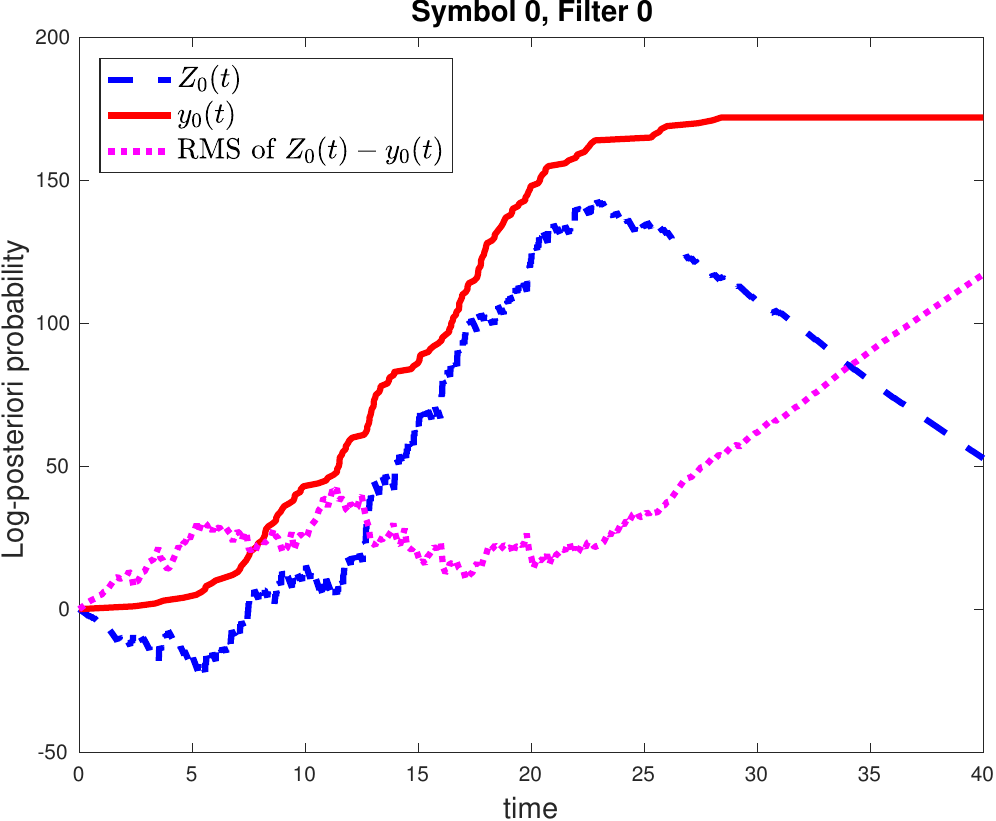}
        \caption{Filter 0 with Symbol 0.}
        \label{fig:diff_1_00}
    \end{subfigure} 
    \begin{subfigure}[t]{0.45\textwidth}
        \centering
        \includegraphics[scale=0.40]{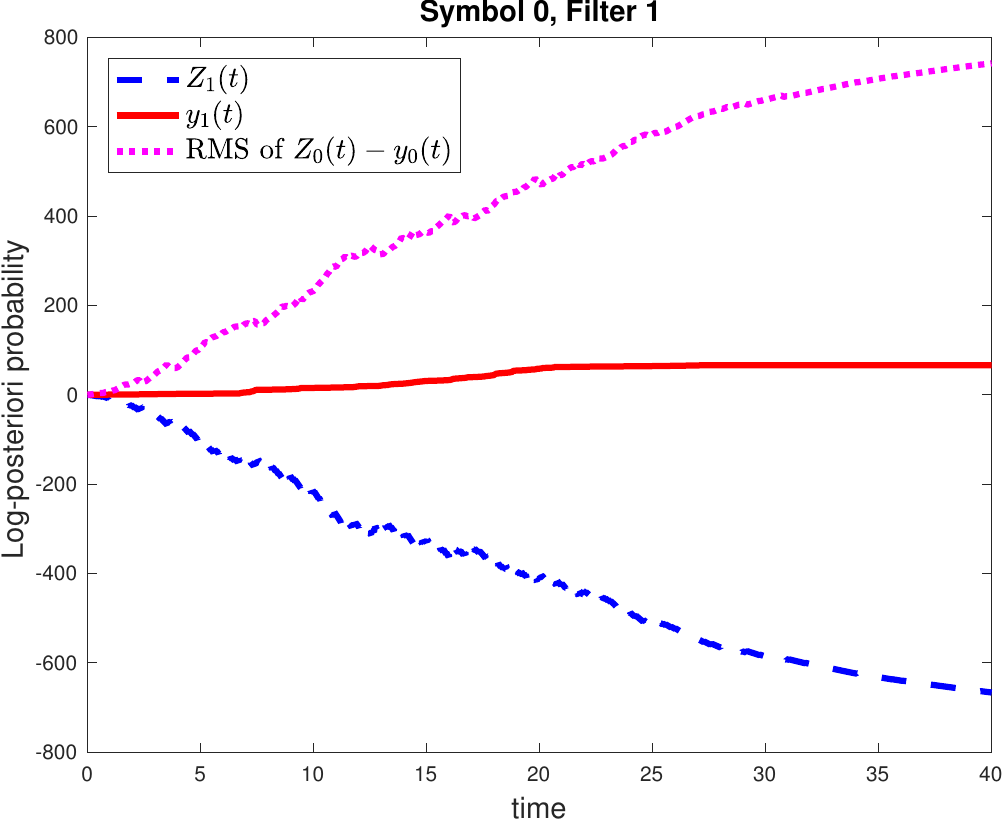}
        \caption{Filter 1 with Symbol 0.}
        \label{fig:diff_1_01}
    \end{subfigure}   
   
    \begin{subfigure}[t]{0.45\textwidth}
        \centering
        \includegraphics[scale=0.40]{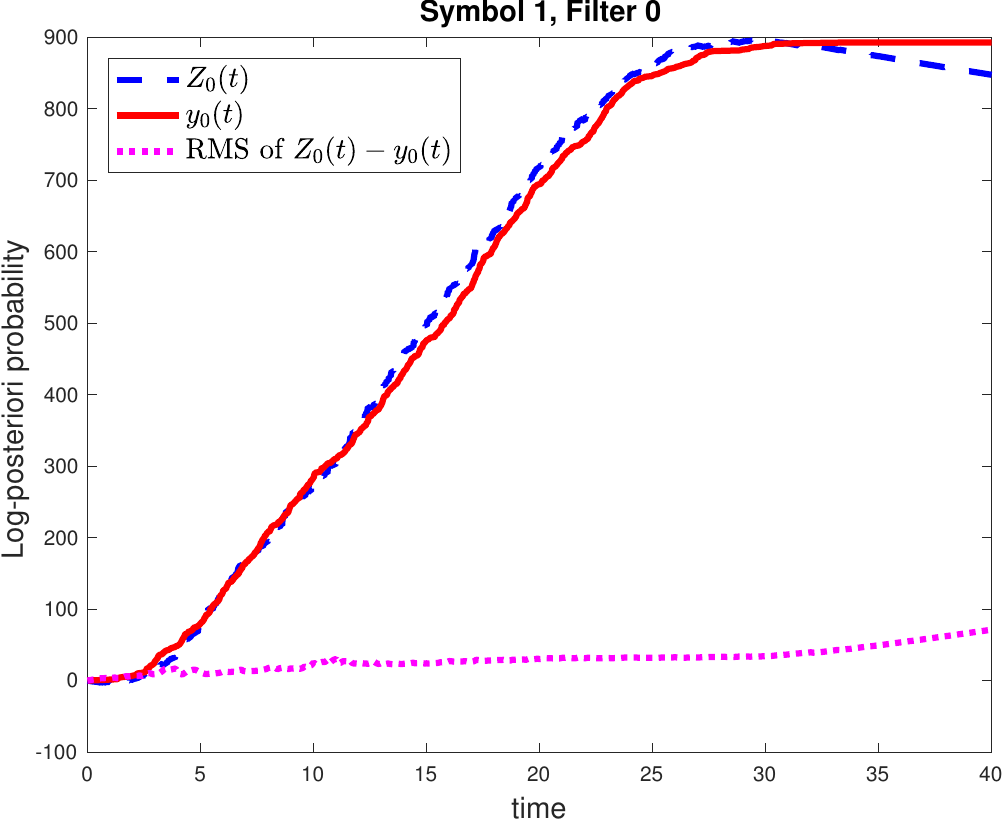}
        \caption{Filter 0 with Symbol 1.}
        \label{fig:diff_1_10}
    \end{subfigure} 
    \begin{subfigure}[t]{0.45\textwidth}
        \centering
        \includegraphics[scale=0.40]{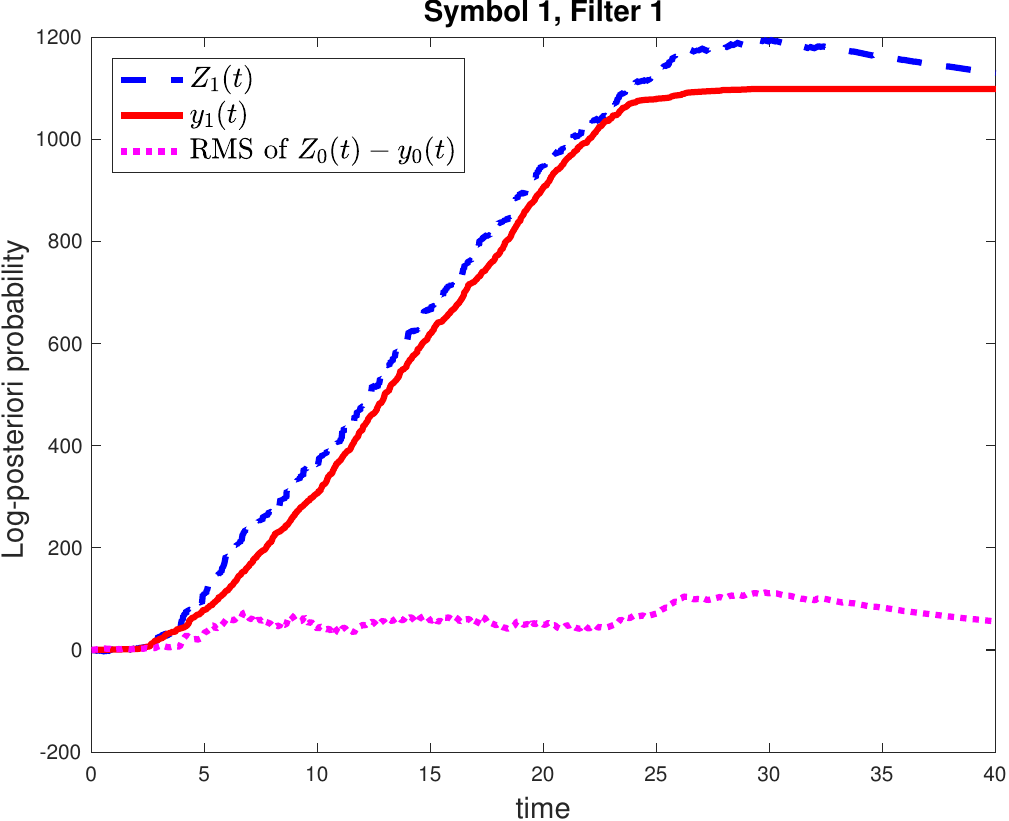}
        \caption{Filter 1 with Symbol 1.}
        \label{fig:diff_1_11}
    \end{subfigure}    
\caption{Comparing molecular demodulation filter $y_k(t)$ against $Z_k(t)$. Note that $Z(t)$ comes from Eq.~\eqref{eqn:logmap_s}.}
\label{fig:diff_1}
\end{figure*}

In order to perform molecular simulation for Eq.~\eqref{eqn:mol_rea}, we assume that the reactions realising the Hill function in Eq.~\eqref{eqn:mol_rea} are fast and reach equilibrium quickly\footnote{\color{black} This is a often used simulation heuristic in biophysics to simulate molecular systems where the Hill function is given and fast. This heuristic can be used to bypass the need to specify the set of elementary chemical reactions that give rise to the Hill function since these reactions are often unknown in biology. For our case, it helps us to bypass the open research problem of selecting the chemical reactions that forms the Hill function, see also Footnote \ref{fn:hill1}. The downside of this heuristic is that noise in the Hill function is ignored. An example of using this heuristic can be found in \cite{Mugler:uk} where a system with fast Michaelis-Menten dynamics (which is similar to the Hill function) is simulated. We leave the problems of designing the elementary reactions for the Hill function and to study its noise dynamics as future work. This future work can build on existing work on stochastic simulation of fast and slow dynamics, see \cite{Schnoerr:2017iw} for a recent survey. 
}. This means we can simulate $y_k(t)$ in Eq.~\eqref{eqn:mol_rea} as an nonhomogenous Poisson process whose instantaneous rate is given by the RHS of Eq.~\eqref{eqn:mol_rea}. We use the method in \cite{Lewis:1979ij} to perform this simulation. Fig.~\ref{fig:diff_1} compares $y_k(t)$ against the approximate log posteriori probability $Z_k(t)$ in Eq.~\eqref{eqn:logmap_s} for $k = 0, 1$ and for Symbols 0 and 1. It can be seen that other than Filter 1 with Symbol 0, we have $y_k(t) \approx Z_k(t)$. The magenta dotted lines show the RMS of $y_k(t) - Z_k(t)$ computed over 100 independent simulations. 

\begin{figure*}[t]
    \centering
    \begin{subfigure}[t]{0.45\textwidth}
        \centering
        \includegraphics[scale=0.40]{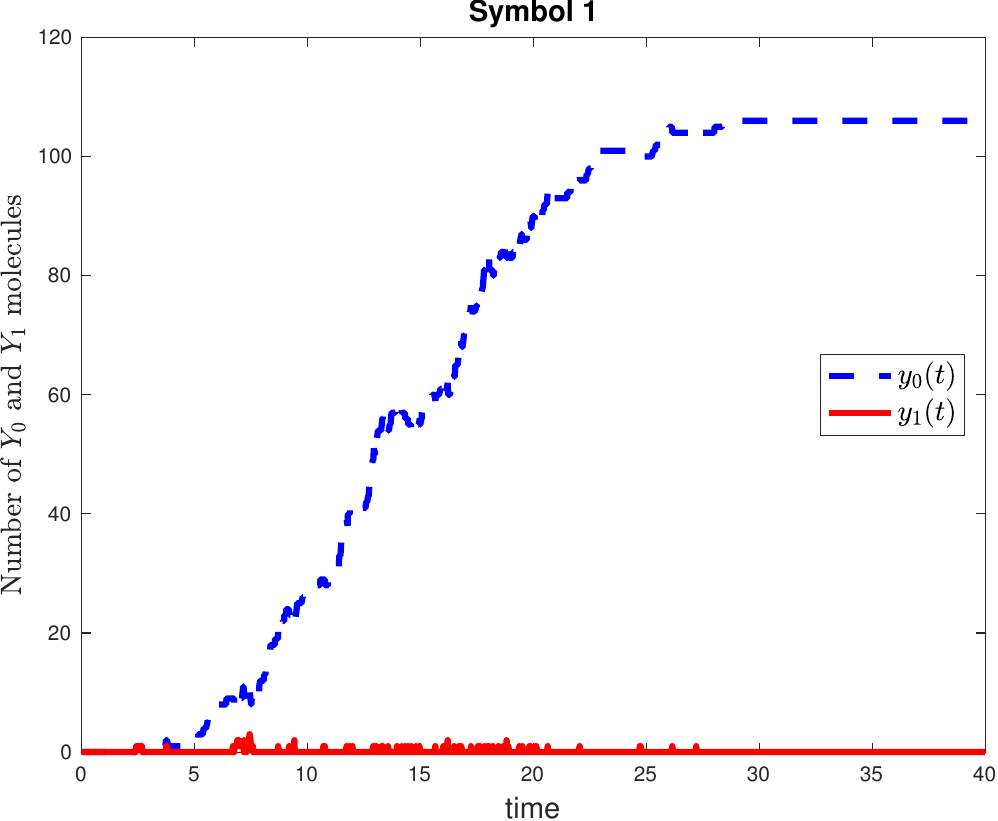}
        \caption{Symbol 0.}
        \label{fig:diff_sub_0}
    \end{subfigure} 
    \begin{subfigure}[t]{0.45\textwidth}
        \centering
        \includegraphics[scale=0.40]{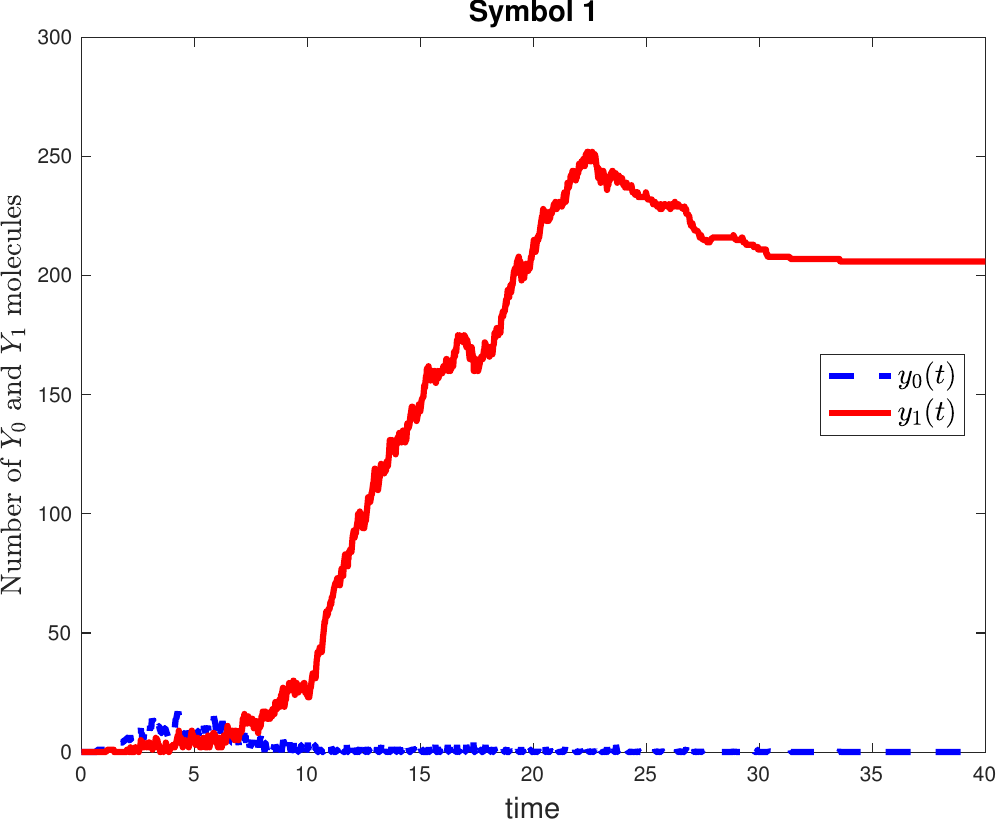}
        \caption{Symbol {\color{black} 1}.}
        \label{fig:diff_sub_1}
    \end{subfigure}      
\caption{$y_0(t)$ and $y_1(t)$ after incorporating annihilation.}
\label{fig:z_anni}
\end{figure*}

The next step is to simulate the annihilation reaction \eqref{cr:anni}. This can be done by using the time series $y_0(k)$ and $y_1(k)$ and use them to simulate a death process. We adapt the method in \cite{Lewis:1979ij} for the simulation of nonhomogenous Poisson process (which we used earlier) for this purpose. We assume $k_a = 1$. Fig.~\ref{fig:z_anni} shows the output after adding the annihilation reaction. It shows that if Symbol $i$ is the input, then species $Y_i$ has a high count most of the time and the other species has a low count because it has been annihilated quickly. 

\begin{figure}[t]
        \centering
	\begin{subfigure}[t]{0.45\textwidth}
        \centering
        \includegraphics[scale=0.40]{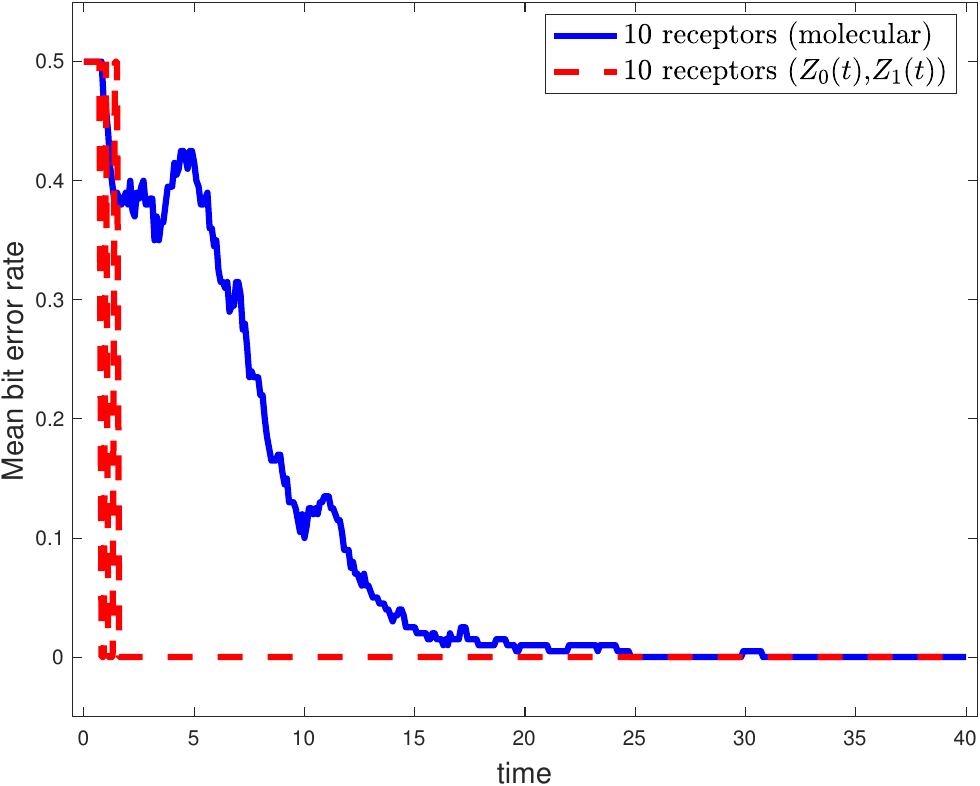}
         \caption{10 receptors.}
        \label{fig:ber_10}
        \end{subfigure}           
        \begin{subfigure}[t]{0.45\textwidth}
        \centering
        \includegraphics[scale=0.40]{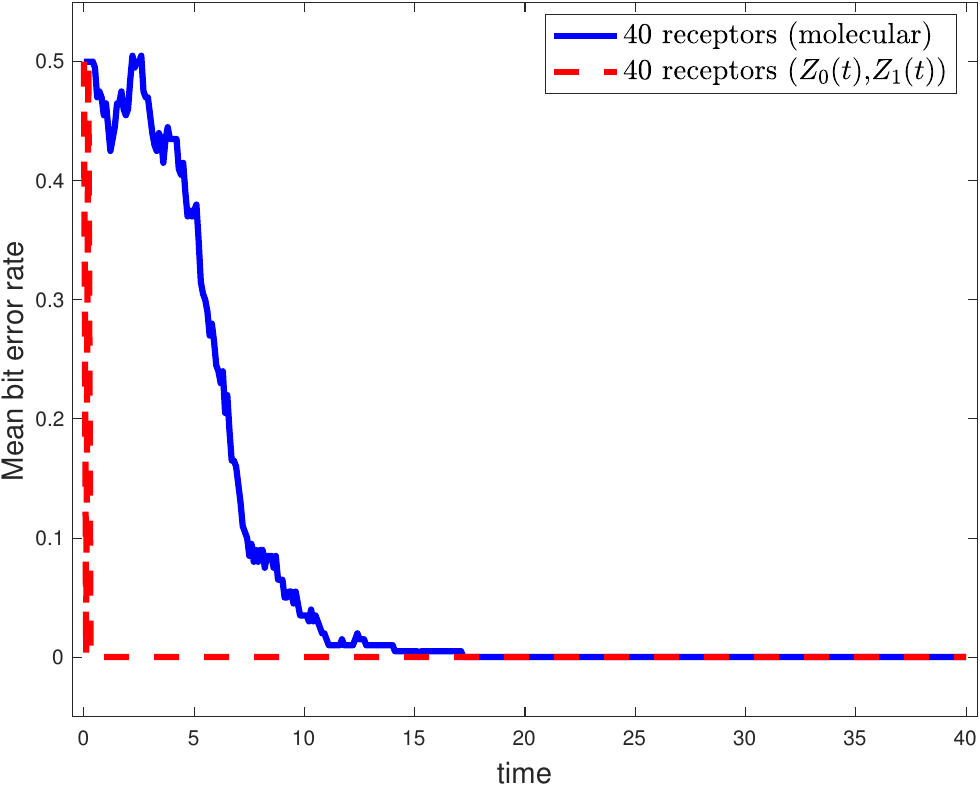}       
         \caption{40 receptors.}        
         \label{fig:ber_40}
        \end{subfigure}          
       \caption{Mean bit error rate for 10 and 40 receptors.}
        \label{fig:ber}
\end{figure}

We now study the impact of the number of receptors on the bit error rate (BER). For a given number of receptors and for each transmitted symbol, we perform 100 independent SSA runs and use the simulation output for the molecular demodulation filters and then the annihilation reaction. The simulation time is 40s. If Symbol $i$ is sent and demodulation decision is made at time $t$, then we say the demodulation is correct if there are more $Y_i$ molecules at time $t$, otherwise it is a bit error. {\color{black} Figs.~\ref{fig:ber_10} and \ref{fig:ber_40} show how the BER varies over time for 10 and 40 receptors. The solid lines are for the molecular demodulation filter while the dashed lines are based on $Z_k(t)$ in Eq.~\eqref{eqn:logmap_s}. Note that the same scale is used in Figs.~\ref{fig:ber_10} and \ref{fig:ber_40}. We observe that for the molecular demodulator, more receptors means BER falls faster with time. Note the response time of the molecular demodulator is slower because there are chemical reaction dynamics in the simulation but these are absent for the demodulator based on $Z(t)$. However, the BER of the molecular demodulator approaches that of $Z(t)$ after some time.  
} 

\begin{figure}[t]
        \centering
        \begin{subfigure}[t]{0.45\textwidth}
        \centering
        \includegraphics[scale=0.40]{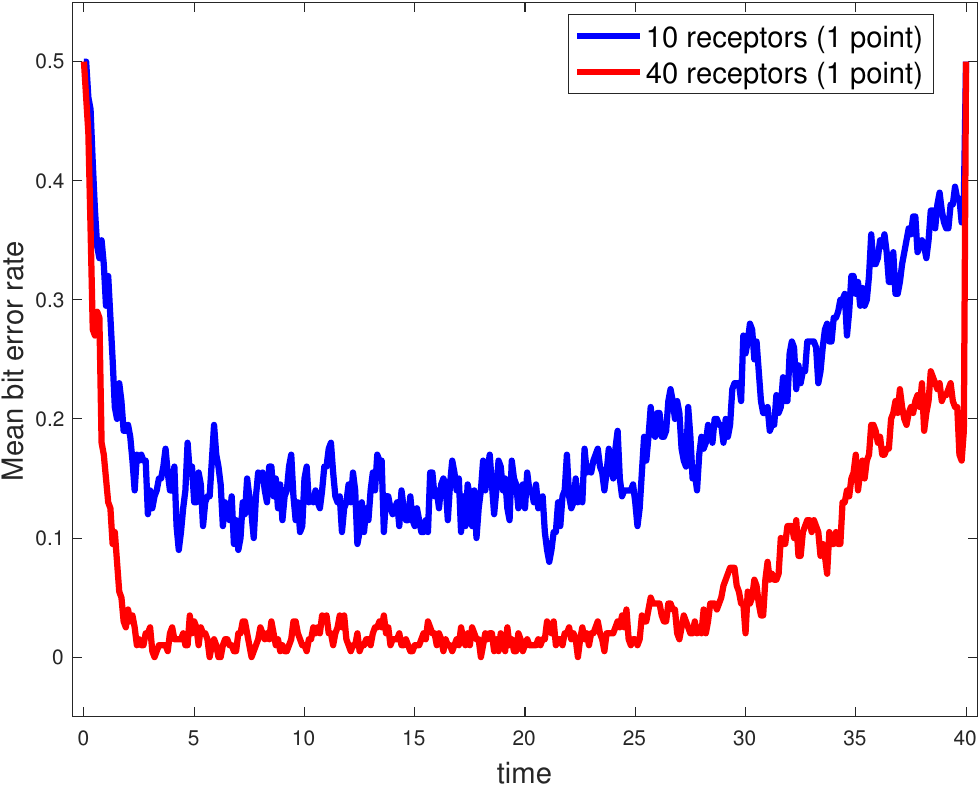}
        \caption{10 and 40 receptors. Time from 0 to 40.}
        \label{fig:ber_1p_a}
        \end{subfigure}           
        \begin{subfigure}[t]{0.45\textwidth}
        \centering
        \includegraphics[scale=0.40]{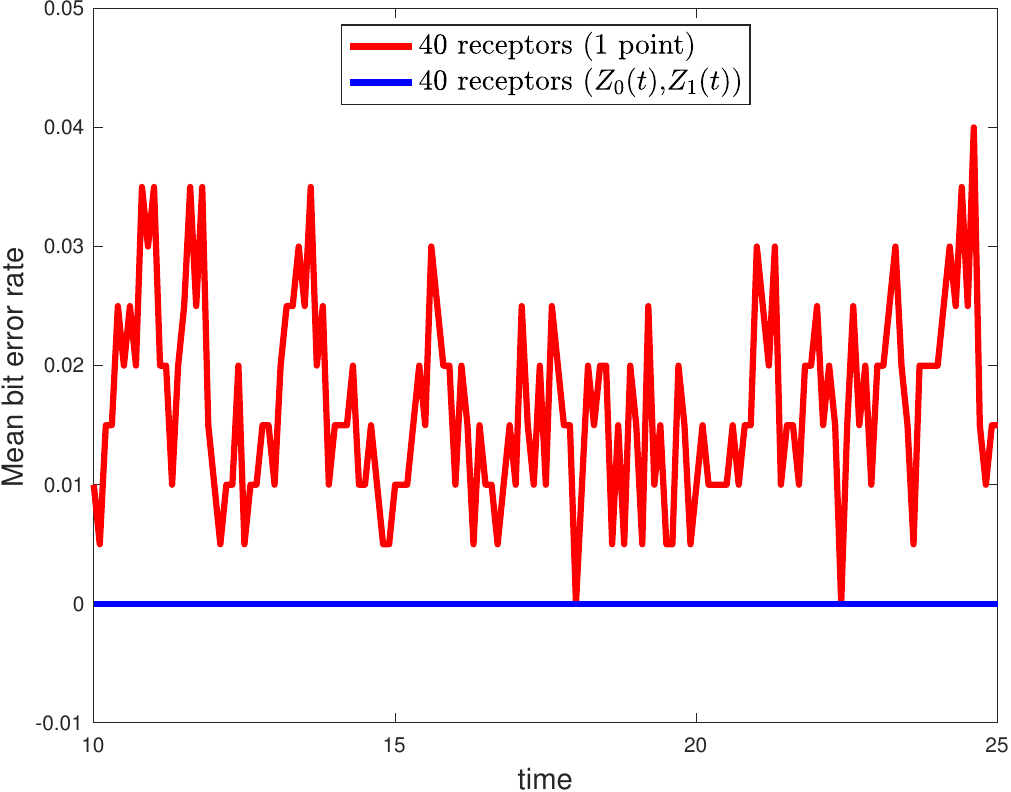}
        \caption{40 receptors. Time from 10 to 25.}
        \label{fig:ber_1p_b}
        \end{subfigure}    
       \caption{Demodulation based on 1 sample of $x_*(t)$.}
        \label{fig:ber_1p}
        
\end{figure}

The demodulator proposed in this paper uses the continuous trajectory of the number of active receptors $x_*(t)$ over time for demodulation. An alternative method of demodulation is to use the number of active receptors at a specific time instant and use a threshold to determine which symbol the transmitter has sent. The aim of this part is to compare these two demodulation methods. We use the same set up as the last paragraph. For demodulation based on one sample point, we determine the BER for this method as follows. If the demodulation decision is to be made at time $t$, then we take the number of active receptors $x_*(t)$ at time $t$ and determine the threshold that gives the least BER for time $t$. We repeat this for many different values of $t$ within the time interval $[0,40]$. {\color{black} Fig.~\ref{fig:ber_1p_a} shows the results for using one sample point for 10 and 40 receptors and Fig.~\ref{fig:ber_1p_b} provides a magnified view for the 40-receptor case in the time interval [10,25]. These figures show that the BER for the 1-sample point method is almost the same in the time interval $[3,20]$. This is because the symbol duration is 20 time units and the reactions in \eqref{cr:all} are fast, therefore the statistical properties of $x_*(t)$ in the time interval $[3,20]$ are almost the same. If we compare the BER in Fig.~\ref{fig:ber_1p_a} against the BER due to $Z_k(t)$ (which uses continuous history) in Figs.~\ref{fig:ber_10} and \ref{fig:ber_40} (Note: all these plots use the same scale.), we see that BER based on continuous history is far lower and is in fact 0 (out of 100 simulation runs) for large enough $t$. This is understandable because continuous trajectory contains full information. The advantage of using history is especially conspicuous for the 10-receptor case where the 1-sample point method gives a BER of around 0.13 but our method gives a much lower BER. This shows that an advantage of using history is to reduce the resource requirement which can be scarce in the nano-bio environment. One may ask whether it is possible to reduce the BER for the 1-sample point method. Two possibilities come to mind. First, one can continue to use one sample point per symbol and use repetition code to send multiple symbols. Second, one can use multiple samples per symbol for demodulation. Our proposed scheme can be viewed as the implementation of these two possibilities. We use a long symbol duration which can be viewed as repetition code. In fact, it is more efficient to use a longer symbol duration than repeating the symbols because we can avoid the transient at the beginning of the symbols. The use of history can be viewed as using multiple samples, or in fact infinite number of samples, in a symbol for demodulation.
}

\begin{remark} There are three parameters that we can use to affect the performance of our demodulator: the pulse amplitudes $a_0$ and $a_1$, and the pulse duration. For example, for a larger separation between transmitter and receiver, one can try to reduce bit error rate by increasing the difference between $a_0$ and $a_1$ or by increasing the pulse duration. 
\end{remark}

\section{Conclusions} 
\label{sec:con} 
This paper presents a method to design a molecular circuit that can demodulate concentration modulated signals in a diffusion-based molecular communication setting. We present numerical experiments to show that the output of the molecular demodulation filter is approximately equal to the positive log-posteriori probability. We also demonstrate that a biochemical circuit from yeast has similar behaviour to the molecular demodulation filter that we have derived. This opens the opportunity to search for natural biochemical circuits that can work as molecular demodulation filters. Although this work focuses on demodulation, the design method presented is also relevant to realising chemical reaction based nano-devices that use analog computation for detection and sensing. 

\subsection*{Acknowledgements}
The author gratefully acknowledges Dr.~Anders Hansen (the lead author of \cite{Hansen:2013fs}) for providing his Matlab code and for answering the author's queries. 


\appendices

\section{Proof} 
\label{sec:no_deriv}
Our aim is to derive the intermediate approximation Eq.~\eqref{eq:Lhat}. We consider a time $t < d$, i.e. when the input signal $u(t)$ is ON. We write $L_k(t)$ in Eq.~\eqref{eqn:logmap_simp} as $L_k(t) = L_{k1}(t) + L_{k2}(t)$ where
\begin{eqnarray}
L_{k1}(t) & = &  \log(a_k) \underbrace{\int_{0}^{t} \left[ \frac{dx_*(t)}{dt} \right]_+ d\tau }_{A(t)}   \label{eq:app_L1} \\
L_{k2}(t) & = & - g_+ a_k \int_{0}^{t} (M - x_*(\tau)) dt. \label{eq:app_L2}
\end{eqnarray}  
and we have used the fact that $\lambda_k(\tau) = a_k, \forall \tau \in [0,d)$. The time series of the number of active receptors $x_*(t)$ is generated by the input signal $u(t)$ where $u(t)$ is one of $K$ possible transmission symbols $\lambda_k(t)$. Recall that we use $a$ to denote the amplitude of the input used where $a$ is one of $a_0$, ..., $a_{K-1}$. 

\ifarxiv
We first consider finding an approximation of the integral $A(t)$ in Eq.~\eqref{eq:app_L1} and the aim is to replace the positive derivative of $x_*(t)$ by some other arithmetic operations which can be computed by using chemical reactions. The integral $A(t)$ can be interpreted as the number of times that the receptors ${\cee X}$ have been activated in the time interval $[0,t]$ when the transmitted symbol is $u(t)$. For an ${\cee X}$ molecule, the time between two consecutive activations is a random variable with mean $m$ and variance $\sigma^2$ where: 
\begin{align}
m =& \frac{1}{g_+ a} + \frac{1}{g_-} \label{eq:app:mean_activation} \\
\sigma^2 = & \frac{1}{(g_+ a)^2} + \frac{1}{g_-^2} 
\end{align}
This is because we can model the activation and deactivation of an ${\cee X}$ molecule by a 2-state continuous-time Markov chain with transition rates $g_+ a$ and $g_-$. 
\fi

We will now make a time-scale separation assumption by assuming that the integration time $t$ in $A(t)$ is much bigger than $m$, i.e. $t  \gg \frac{1}{g_+ a} + \frac{1}{g_-}$. This assumption can be met by having a sufficiently long integration time $t$ and large amplitude $a$. If this time-scale separation assumption holds, then there are many activations in the time interval $[0,t]$. In this case, we can use the renewal theorem \cite{Grimmett} to approximate $A(t)$, we have:
${\rm mean}(A(t))  \approx  M \frac{t}{m}$ and  
${\rm var}(A(t))  \approx M \frac{\sigma^2}{m^3} t$,    
which implies that  
$\frac{ \sqrt{ {\rm var}(A(t)) } }{ {\rm mean}(A(t)) } \approx  \frac{ \sigma }{  m \sqrt{M} \sqrt{t} }.$
This means we can approximate $A(t)$ by its mean and the error decreases with the reciprocal of the square root of the integration time $t$. By using this approximation, we have:
\begin{eqnarray}
L_{k1}(t) & \approx & \log(a_k) \frac{M}{m} t \label{eq:app_L1_1} 
\end{eqnarray}  

The time-scale separation assumption also implies that the continuous-time Markov chain describing the number of ${\cee X_*}$ reaches equilibrium quickly. Therefore,  the ensemble average of $x_*(t)$ can be treated as a constant in the time interval $[0,t]$; we will denote this average by $x_{*,a}$ where 
\begin{eqnarray}
x_{*,a} & = & \frac{M g_+ a}{g_+ a + g_-} \label{eq:app:ensemble_average} 
\end{eqnarray}

\ifarxiv
This ensemble average is related to mean inter-activation time $m$ in Eq.~\eqref{eq:app:mean_activation} by:
\begin{eqnarray}
x_{*,a} & = & \frac{M}{m g_-}
\end{eqnarray}
\fi
By using this relationship in Eq.~\eqref{eq:app_L1_1}, we have:
\begin{eqnarray}
L_{k1}(t) & \approx & \log(a_k) \; g_- x_{*,a} \;  t .    \label{eq:app_L1_2} 
\end{eqnarray}  
We will return to this expression shortly after studying the approximation of the integral in $L_{k2}(t)$ in Eq.~\eqref{eq:app_L2}. 

Since the Markov chain describing the reaction cycle of ${\cee X}$ and ${\cee X_*}$ is ergodic, the time average in Eq.~\eqref{eq:app_L2} can be approximated by its ensemble average. We have 
\begin{eqnarray}
L_{k2}(t) & \approx & - g_+  a_k \;  (M - x_{*,a})  \; t .  \label{eq:app_L2_2_inter} 
\end{eqnarray}  
The next step is to replace the $x_{*,a}$ in Eq.~\eqref{eq:app_L2_2_inter} by the RHS of Eq.~\eqref{eq:app:ensemble_average} to arrive at: 
\begin{eqnarray}
L_{k2}(t) & \approx & - a_k \; \frac{g_- x_{*,a}}{a} \; t  \label{eq:app_L2_2} 
\end{eqnarray}    
  
Since $L_k(t) = L_{k1}(t) + L_{k2}(t)$, it follows from Eqs.~\eqref{eq:app_L1_2} and \eqref{eq:app_L2_2} that:
\begin{eqnarray}
L_k(t) & \approx & g_-  \; x_{*,a}  \left( \log(a_k)  - \frac{a_k}{a}    \right)  t     \label{eq:app_L_3} 
\end{eqnarray}  
for $t \leq d$. 

We can re-write Eq.~\eqref{eq:app_L_3} in differential form, as follows: 
\begin{eqnarray}
\frac{dL_k(t)}{dt} & \approx &  g_-  \; x_*(t) \left\{ \log(\lambda_k(t) )  -  \frac{\lambda_k(t)}{u(t)} \right\} \label{eq:star:L2} 
\end{eqnarray} 

\section{Maximum mYFP} 
\label{app:mYFP}
In this Appendix, we will explain why, for a given input Msn2 amplitude, the maximum mYFP in the CM-inspired model is proportional to the total duration that Msn2 is ON. We first state the complete CM-inspired model:
\begin{align}
\frac{d P_{\rm active}(t)}{dt}  =  & g_+ [Msn2](t) \; (1 - P_{\rm active}(t)) - g_- \; P_{\rm active}(t) \label{eq:app:cm_bio_1} \\
\frac{d [C_{\rm init}](t)}{dt}  = & g_- \; P_{\rm active}(t) \times  \frac{h [Msn2](t)^{n}}{H^{n} + [Msn2](t)^{n}}   - d_2 C_{\rm init} \label{eq:app:cm_bio_2}  
\end{align}
\begin{align}
\frac{d [mRNA](t)}{dt} = & k_3 [C_{\rm init}](t) - d_3 [mRNA](t) \label{eq:app:cm_bio_3} \\
\frac{d [YFP](t)}{dt} = & k_4 [mRNA](t) - (d_4 + k_5) [YFP](t) \label{eq:app:cm_bio_4} \\
\frac{d [mYFP](t)}{dt} = & k_5 [YFP](t) - d_4 [mYFP](t) \label{eq:app:cm_bio_5} 
\end{align} 
where $k_4$, $d_4$, $k_5$ are reaction rate constants. 

From \cite{Hansen:2013fs}, we know that the degradation rate $d_4$ of mYFP is small. We will therefore assume $d_4 = 0$ in Eq.~\eqref{eq:app:cm_bio_5}. This means the concentration of mYFP is monotonically increasing and we can see from Fig.~\ref{fig:bio_1} that this is approximately true. As a result, the maximum concentration of mYFP, denoted by $[mYFP]_{\max}$, can be approximated by $[mYFP](t)$ at $t = \infty$. By using Eq.~\eqref{eq:app:cm_bio_5}, we have 
\begin{align}
[mYFP]_{\max} = \int_{0}^\infty k_5 \; [YFP](t) \; dt 
\end{align} 

We now state a result that was proved in Supplementary Figure 11b in \cite{Hao:2011kz}. Consider the ODE:
\begin{align}
\frac{d \xi(t)}{dt}  =  & \alpha \; \chi(t) - \beta \; \xi(t) 
\end{align}
where $\chi(t)$ is a bounded function that vanishes at $t = \infty$, i.e. $\chi(\infty) = 0$. It was shown that:
\begin{align}
\int_{0}^\infty \xi(t) dt =  & \frac{\alpha}{\beta}  \int_{0}^\infty \chi(t) dt 
\end{align}

By sequentially applying this result to Eqs.~\eqref{eq:app:cm_bio_4}, \eqref{eq:app:cm_bio_3} and \eqref{eq:app:cm_bio_2}, we can show that 
\begin{align}
[mYFP]_{\max} = \frac{k_4 k_3 g_-}{d_3  d_2} \int_{0}^\infty P_{\rm active}(t)  \frac{h [Msn2](t)^{n}}{H^{n} + [Msn2](t)^{n}} dt 
\label{eq:app:mYFP_int}
\end{align} 
Since the Hill function coefficients are chosen so that Eq.~\eqref{eq:cm_bio_2_hill} holds, therefore the Hill function is approximately 
\begin{eqnarray}
\left[\log(a) - \frac{a}{[Msn2](t)} \right]_+ 
\label{eq:app:mYFP_matching} 
\end{eqnarray}
Let us now assume that $[Msn2](t)$ is a piecewise constant signal with 2 possible levels, which we will call ON and OFF. When the signal is OFF, its amplitude is small and result in the expression in Eq.~\eqref{eq:app:mYFP_matching} being zero. When the signal is ON, its amplitude is large enough and result in the expression in Eq.~\eqref{eq:app:mYFP_matching} being positive and constant. In addition, if the reactions \eqref{cr:all} are fast relative to duration of the pulses, then $P_{\rm active}(t)$ can be considered to be a constant. Therefore, by using these properties in the integral in Eq.~\eqref{eq:app:mYFP_int}, it can be seen that, for a given $[Msn2](t)$ ON-amplitude, $[mYFP]_{\max}$ is approximately proportional to the time when $[Msn2](t)$ is ON. 

\ifarxiv
\section{Problem for the $K \geq 3$ case} 
\label{app:anni}
We will explain the problem using $K = 3$. Consider three chemical species $\cee{Y_k}$ which are the output of the molecular demodulation filters $y_k(t)$. Let us assume that the three species annihilate each other with reactions 
\begin{align}
\cee{Y_i + Y_j ->[k_a] \phi }
\end{align} 
where $i \neq j$. Let $\rho_k(t)$ be the production rate of $\cee{Y_k}$ at time $t$. A differential equation model for the concentration $[Y_k](t)$ $(k = 0, 1, 2)$ is given by:
\begin{align}
\frac{d [Y_0](t)}{dt} = & \rho_0(t) - k_a  [Y_1](t) \;  [Y_2](t) \\
\frac{d [Y_1](t)}{dt} = & \rho_1(t) - k_a  [Y_0](t) \;  [Y_1](t) \\
\frac{d [Y_2](t)}{dt} = & \rho_2(t) - k_a  [Y_0](t) \;  [Y_2](t) 
\end{align}
The difficulty is that the steady state of this set of differential equations depends on the dynamics $\rho_k(t)$. We will illustrate this using a numerical example. Let $\delta(t)$ denote the Dirac delta function. Consider the following two sets of $\rho_k(t)$.
\begin{itemize}
\item $\rho_0(t) = 20 \delta(t)$, $\rho_1(t) = 30 \delta(t)$, $\rho_2(t) = 40 \delta(t-10)$
\item $\rho_0(t) = 20 \delta(t)$, $\rho_1(t) = 30 \delta(t-10)$, $\rho_2(t) = 40 \delta(t)$
\end{itemize}
Note that for both sets of $\rho_k(t)$, the number of ${\cee Y_k}$ molecules generated is the same for each $k$; however the times at which they are generated are different. We assume $k_a \rightarrow \infty$ so reactions are very fast. 

Consider the first case. There are only ${\cee Y_0}$ and ${\cee Y_1}$ in the beginning so they annihilate each other. This means there are 10 ${\cee Y_1}$ molecules left before 40 ${\cee Y_2}$ are released at $t = 10$. At steady state, there are 30 ${\cee Y_2}$ molecules. 

Now consider the second case. There are only ${\cee Y_0}$ and ${\cee Y_2}$ in the beginning so they annihilate each other. This means there are 20 ${\cee Y_2}$ molecules left before 30 ${\cee Y_1}$ are released at $t = 10$. At steady state, there are 10 ${\cee Y_1}$ molecules. 

Ideally, we would always like to have only the chemical species with the highest count left. However, the example above demonstrates that for $K = 3$, this is not always the case. Although in some chemical reaction systems, the stochastic dynamics can be different from the deterministic dynamics, our observation based on simulation shows that the stochastic and deterministic dynamics for this system are the same. 
\fi

\end{document}